\batchmode

\documentclass[12pt,epsf]{article}

\usepackage{times}
\usepackage{graphicx}
\usepackage{amsfonts}
\usepackage{amsmath}
\usepackage{amssymb}
\usepackage{amstext}
\usepackage{amsthm}
\usepackage{youngtab}

\usepackage{epsfig}
\usepackage{color}
\usepackage{leftidx}
\usepackage{tikz}
\psfigdriver{dvips}
\usepackage{latexsym}
\usepackage[matrix,curve]{xypic}
\usepackage{rotating}
\rotdriver{dvips}

\begin{document}

\baselineskip 6mm
\renewcommand{\thefootnote}{\fnsymbol{footnote}}


\newcommand{\nc}{\newcommand}
\newcommand{\rnc}{\renewcommand}


\rnc{\baselinestretch}{1.24}    
\setlength{\jot}{6pt}       
\rnc{\arraystretch}{1.24}   

\makeatletter
\rnc{\theequation}{\thesection.\arabic{equation}}
\@addtoreset{equation}{section}
\makeatother



\nc{\be}{\begin{equation}}

\nc{\ee}{\end{equation}}

\nc{\bea}{\begin{eqnarray}}

\nc{\eea}{\end{eqnarray}}

\nc{\xx}{\nonumber\\}

\nc{\ct}{\cite}

\nc{\la}{\label}

\nc{\eq}[1]{(\ref{#1})}

\nc{\newcaption}[1]{\centerline{\parbox{6in}{\caption{#1}}}}

\nc{\fig}[3]{

\begin{figure}
\centerline{\epsfxsize=#1\epsfbox{#2.eps}}
\newcaption{#3. \label{#2}}
\end{figure}
}


\def\CA{{\cal A}}
\def\CC{{\cal C}}
\def\CD{{\cal D}}
\def\CE{{\cal E}}
\def\CF{{\cal F}}
\def\CG{{\cal G}}
\def\CH{{\cal H}}
\def\CK{{\cal K}}
\def\CL{{\cal L}}
\def\CM{{\cal M}}
\def\CN{{\cal N}}
\def\CO{{\cal O}}
\def\CP{{\cal P}}
\def\CR{{\cal R}}
\def\CS{{\cal S}}
\def\CU{{\cal U}}
\def\CV{{\cal V}}
\def\CW{{\cal W}}
\def\CY{{\cal Y}}
\def\CZ{{\cal Z}}


\def\IB{{\hbox{{\rm I}\kern-.2em\hbox{\rm B}}}}
\def\IC{\,\,{\hbox{{\rm I}\kern-.50em\hbox{\bf C}}}}
\def\ID{{\hbox{{\rm I}\kern-.2em\hbox{\rm D}}}}
\def\IF{{\hbox{{\rm I}\kern-.2em\hbox{\rm F}}}}
\def\IH{{\hbox{{\rm I}\kern-.2em\hbox{\rm H}}}}
\def\IN{{\hbox{{\rm I}\kern-.2em\hbox{\rm N}}}}
\def\IP{{\hbox{{\rm I}\kern-.2em\hbox{\rm P}}}}
\def\IR{{\hbox{{\rm I}\kern-.2em\hbox{\rm R}}}}
\def\IZ{{\hbox{{\rm Z}\kern-.4em\hbox{\rm Z}}}}


\def\a{\alpha}
\def\b{\beta}
\def\d{\delta}
\def\ep{\epsilon}
\def\ga{\gamma}
\def\k{\kappa}
\def\l{\lambda}
\def\s{\sigma}
\def\t{\theta}
\def\w{\omega}
\def\G{\Gamma}


\def\half{\frac{1}{2}}
\def\dint#1#2{\int\limits_{#1}^{#2}}
\def\goto{\rightarrow}
\def\para{\parallel}
\def\brac#1{\langle #1 \rangle}
\def\curl{\nabla\times}
\def\div{\nabla\cdot}
\def\p{\partial}


\def\Tr{{\rm Tr}\,}
\def\det{{\rm det}}


\def\vare{\varepsilon}
\def\zbar{\bar{z}}
\def\wbar{\bar{w}}
\def\what#1{\widehat{#1}}


\def\ad{\dot{a}}
\def\bd{\dot{b}}
\def\cd{\dot{c}}
\def\dd{\dot{d}}
\def\so{SO(4)}
\def\bfr{{\bf R}}
\def\bfc{{\bf C}}
\def\bfz{{\bf Z}}

\begin{titlepage}


\hfill\parbox{3.7cm} {{\tt arXiv:2109.00001}}

\vspace{15mm}

\begin{center}
{\Large \bf  Anatomy of Einstein Manifolds}

\vspace{10mm}

Jongmin Park \footnote{whdalsdl776@gist.ac.kr}, Jaewon Shin \footnote{j1thegreat@gm.gist.ac.kr}
and Hyun Seok Yang \footnote{hsyang@gist.ac.kr}
\\[10mm]

{\sl Department of Physics and Photon Science \\ Gwangju Institute of Science and Technology, Gwangju 61005, Korea}

\end{center}

\thispagestyle{empty}

\vskip1cm


\centerline{\bf ABSTRACT}
\vskip 4mm
\noindent

An Einstein manifold in four dimensions has some configuration of $SU(2)_+$ Yang-Mills instantons and
$SU(2)_-$ anti-instantons associated with it.
This fact is based on the fundamental theorems that the four-dimensional Lorentz group $Spin(4)$
is a direct product of two groups $SU(2)_\pm$ and the vector space of two-forms decomposes
into the space of self-dual and anti-self-dual two-forms.
It explains why the four-dimensional spacetime is special for the stability of Einstein manifolds.
We now consider whether such a stability of four-dimensional Einstein manifolds
can be lifted to a five-dimensional Einstein manifold.
The higher-dimensional embedding of four-manifolds from the viewpoint of gauge theory
is similar to the grand unification of Standard Model
since the group $SO(4) \cong Spin(4)/\mathbb{Z}_2 = SU(2)_+ \otimes SU(2)_-/\mathbb{Z}_2$
must be embedded into the simple group $SO(5) = Sp(2)/\mathbb{Z}_2$.
Our group-theoretic approach reveals the anatomy of Riemannian manifolds
quite similar to the quark model of hadrons in which two independent Yang-Mills instantons
represent a substructure of Einstein manifolds.
\\



\vspace{1cm}

\today

\end{titlepage}

\renewcommand{\thefootnote}{\arabic{footnote}}
\setcounter{footnote}{0}

\section{Introduction}

The hadrons we know all fall into multiplets that reflect underlying internal symmetries.
To express this fact in a simple and concrete way, it was hypothesized that hadrons are composed of
more elementary constituents with basic symmetries, called quarks.
The $SU(3)$ multiplet structure of the hadrons (baryons and mesons) strongly hinted
at the existence of a substructure \cite{gell-mann1962,penn4}. According to the quark model \cite{gell-mann1964},
all hadrons are made up of quarks and anti-quarks, bound together in different ways.
Even in the absence of knowledge about the potential which binds quarks and anti-quarks,
the model was very predictive. The triple tensor product of the fundamental representation $\mathbf{3}$ of
the $SU(3)$ flavor symmetry leads to octets and a decuplet of baryons, $\mathbf{3} \otimes \mathbf{3} \otimes
\mathbf{3} = \mathbf{1} \oplus \mathbf{8} \oplus \mathbf{8} \oplus \mathbf{10}$, in addition to a singlet.
This classification works also for mesons, $\mathbf{3} \otimes \overline{\mathbf{3}}
= \mathbf{1} \oplus \mathbf{8}$.
This quark model eventually led to the introduction of color degrees of freedom and the construction
of quantum chromodynamics \cite{color1,color2}.

A special feature, which permeates four-dimensional geometry, is the fact that $Spin(4)$ splits
into a product of two groups:
\begin{equation}\label{spin4-split}
  Spin(4) = SU(2)_+ \times SU(2)_-.
\end{equation}
The group $Spin(4)$ is a double cover of the four-dimensional Euclidean Lorentz group $SO(4)$,
i.e. $SO(4)\cong SU(2)_+ \times SU(2)_-/\mathbb{Z}_2$.
The splitting of $Spin(4)$ is isomorphically related to the decomposition of the 2-forms on a four-manifold.
Using the Hodge $*$-operator acting on exterior 2-forms, one can split 2-forms into self-dual and anti-self-dual 2-forms.
The splitting can be applied to the curvature form of any bundle with connection over an oriented four-manifold.
The canonical splitting of the vector spaces leads to the irreducible decomposition of
Riemann curvature tensor $R \in C^\infty ( \mathfrak{g} \otimes \Omega^2)$ as \cite{s-duality}
\begin{equation}\label{riem-irrdec}
  R = R_{(++)} \oplus  R_{(+-)} \oplus R_{(-+)} \oplus R_{(--)}
\end{equation}
where the subscript $(\pm \pm)$ refers to the splitting of the vector spaces
$\mathfrak{g} \equiv so(4) = su(2)_+ \oplus su(2)_-$ and $\Omega^2 \equiv \Lambda^2 T^* M = \Omega^2_+ \oplus \Omega^2_-$.
This splitting of the vector spaces occupies a central position for the Donaldson theory of four-manifolds and
has been well-known in mathematical literatures (see, for example, Chaps. 1.G \& 1.H in \cite{book-besse} and
Sects. 1.1 \& 2.1 in \cite{book-dokr}).

Imposing the Einstein equations, $R_{\mu\nu} = \lambda g_{\mu\nu}$, leads to the condition
$R_{(+-)} = R_{(-+)}^T = 0$ (\textbf{6.32} in \cite{book-besse} and Lemma in \cite{oh-yang}).
In this case, the Riemann curvature tensor satisfies the self-duality equations,
$* R_{(\pm \pm)} =  R_{(\pm \pm)} * = \pm  R_{(\pm \pm)}$, where
the Hodge $*$-operator $* R$ acts on the first two indices $[ab]$ of the curvature tensor $R_{abcd}$ and
$R*$ acts on the last two indices $[cd]$.
Therefore, an Einstein manifold consists of $SU(2)_+$ instantons and $SU(2)_-$ anti-instantons
defined over itself \cite{oh-yang,joy-jhep}.
The instantons in $SU(2)_+$ group live in a different representation space from the anti-instantons in $SU(2)_-$ group
because these are two independent factors for the product group \eq{spin4-split}.
This means that $R_{(++)}$ and $R_{(--)}$ correspond to two independent components defined by self-dual and
anti-self-dual spin connections, respectively, acting on the chiral and anti-chiral spin bundles.
This special feature of four-dimensional gravity has been originated from the splitting of
the product group \eq{spin4-split}.

A four-dimensional Einstein manifold has the irreducible decomposition defined by the curvature tensor
$R = R_{(++)} \oplus R_{(--)}$ which brings about two independent gravitational components.
However, such division into two independent instanton sectors explains the stability of Einstein manifolds.
It turns out \cite{pos-yang} that the topological invariants carried by an Einstein manifold
are determined by the configuration of $SU(2)_+$ instantons and $SU(2)_-$ anti-instantons,
as will be reviewed in Sec. 2.
Therefore, an Einstein manifold has a substructure like hadrons.
An interesting physics arises if the four-dimensional gravity is regarded as being obtained
from a five-dimensional gravity through the Kaluza-Klein compactification \cite{kk-book}.
The Riemann curvature tensor in five dimensions takes values in the Lie algebra of
the Lorentz group $SO(5)$. The group $SO(5)$ is a simple group
unlike the four-dimensional Lorentz group $SO(4)= SU(2)_+ \times SU(2)_-/\mathbb{Z}_2$.
Since the group $SO(4)$ must be embedded into the simple group $SO(5) = Sp(2)/\mathbb{Z}_2$
in the five-dimensional gravity, we expect that two independent components
caused by the separation of Riemann curvature tensors will be combined into a single gravitation force
in five dimensions. Moreover, the electromagnetism and a scalar field obtained
from a five-dimensional metric through the Kaluza-Klein reduction
should also appear in the same multiplet in an irreducible representation (irrep) of
the Lorentz group $SO(5)$.

This unification scheme is similar to the grand unification of Standard Model
since the group $SO(4) \cong SU(2)_+ \otimes SU(2)_-/\mathbb{Z}_2$ must be embedded
into the simple group $SO(5) = Sp(2)/\mathbb{Z}_2$ although the Kaluza-Klein theory
is reduced from a five-dimensional gravity.
The Standard Model has a product gauge group $SU(3) \times SU(2) \times U(1)$ to describe
the electroweak and strong forces.
In the grand unified theory (GUT), the product gauge group in Standard Model is embedded into a single gauge group,
for example, $SU(5)$ or $SO(10)$ (see, e.g., Chaps. 18 \& 24 in \cite{georgi-book}).
The leptons and quarks in the GUT appear in the same multiplet in a larger symmetry.
The unification of forces with a larger simple group typically opens a new decay channel
of proton into leptons and so introduces a novel instability of a stable particle in Standard Model.
We will see how two instanton sectors of four-dimensional Einstein manifolds are similarly combined into
a five-dimensional Einstein manifold. The embedding of $SU(2)_+$ instantons and $SU(2)_-$ anti-instantons
into a five-dimensional Einstein manifold may similarly develop a novel instability like the proton decay
in the GUT. A speculative reason for this assumption is that
there is no natural topological invariant such as the Euler characteristic
or the Hirzebruch signature in five dimensions \cite{book-besse,egh-report} that
guarantees the stability of five-dimensional Einstein manifolds.

The motivation for the present work lies in providing a fresh point of view for the topological structure
of Einstein manifolds using the group properties of $SO(4) \cong SU(2)_+ \otimes SU(2)_-/\mathbb{Z}_2$ and
$SO(5) = Sp(2)/\mathbb{Z}_2$.
We hope it will provide a deeper insight into the nature of the stability of Riemannian manifolds.
We will mostly refer to the perturbative stability of Einstein manifolds regarding to the second variation of 
the Einstein-Hilbert action with a fixed volume at a background Einstein metric (see Chaps. 4 \& 12 in \cite{book-besse}).
But a nonperturbative instability may be induced by instanton transitions.
We will not try to exhaust all the details but initiate a work along this direction.

This paper is organized as follows. In section 2, we briefly review how the decomposition of
Riemann curvature tensor \eq{riem-irrdec} is derived from the splitting of the vector spaces
$\mathfrak{g} \equiv so(4) = su(2)_+ \oplus su(2)_-$ and $\Omega^2 \equiv \Lambda^2 T^* M = \Omega^2_+ \oplus \Omega^2_-$.
We also discuss how topological invariants of Einstein manifolds such as
the Euler characteristic and the Hirzebruch signature are determined by the configuration
of $SU(2)_+$ instantons and $SU(2)_-$ anti-instantons to illuminate a substructure of Einstein manifolds.
In section 3, we consider a five-dimensional Einstein manifold and its Kaluza-Klein reduction.
We expand the five-dimensional Riemann curvature tensor in the basis of $sp(2) \cong so(5)$ Lie algebra
which generalizes the decomposition \eq{riem-irrdec} to five dimensions.
After the Kaluza-Klein reduction, the Lorentz symmetry $SO(5)$ is spontaneously
broken to $SO(4) \times U(1)$ where $U(1)$ is originated from the isometries of the Kaluza-Klein circle \cite{kk-book}.
According to the symmetry breaking pattern, we further decompose the five-dimensional Riemann curvature tensor
in the basis of $so(4) \cong su(2)_+ \oplus su(2)_-$ Lie algebra.
This decomposition is useful to see how $U(1)$ gauge fields and a scalar field deform the instanton structure
of four-dimensional Einstein manifolds and to understand how these deformed geometries are nicely combined
into a five-dimensional Einstein manifold.
In section 4, we consider particular cases to consolidate that all these deformations can be organized
into a single five-dimensional Einstein manifold once the fifth dimension is opened so that
the Lorentz symmetry is enhanced to $SO(5)$.
In section 5, we discuss some important issues and generalization to non-compact Einstein manifolds
that we have not addressed in this paper, and speculate a possible origin
of novel instabilities of Einstein manifolds in five dimensions.

In Appendix A, we provide some details about the Lie algebras $so(4) \cong su(2)_+ \oplus su(2)_-$ and
$sp(2) \cong so(5)$. The geometric details of five-dimensional gravity and Kaluza-Klein gravity,
especially in the vielbein formalism, appear in Appendix B.
Appendix C contains the group structure analysis of Riemann curvature tensors and the decomposition of
Ricci tensors and Ricci scalar in the $so(4)$ Lie algebra basis.

\section{Einstein manifolds as Yang-Mills instantons}

It is known \cite{eins-stable1,eins-stable2} that Einstein manifolds are stable at least perturbatively.
It is a bit mysterious, recalling that gravity is also described within the framework of field theory.
One way to understand the stability is to notice that an Einstein manifold carries nontrivial topological invariants
such as the Euler characteristic $\chi$ and Hirzebruch signature $\tau$ \cite{egh-report}.
The gauge theory formulation of gravity reveals a beautiful aspect of the stability.
It turns out \cite{oh-yang,joy-jhep} that an Einstein manifold in four dimensions has a configuration
of $SU(2)_+$ Yang-Mills instantons and $SU(2)_-$ anti-instantons.
Two kinds of instantons are independent of each other because they belong
to different gauge groups.\footnote{This reasoning may not be complete because a new instability may be
developed through the interaction between instantons.
Moreover, $\mathbb{T}^4$ and $\mathbb{S}^1 \times \mathbb{S}^3$ (which is not an Einstein manifold)
have trivial topological invariants.
However, the stability of these product manifolds may be guaranteed by a lower-dimensional topology.}
Furthermore, instantons can be superposed to make multi-instantons.
In principle, it is possible to have a tower of Einstein manifolds by superposing $SU(2)$ instantons
in each gauge group. The multi-Taub-NUT spaces \cite{euc-bh} could serve as an example of such tower
(with only one type of instantons used).
Of course, a compact manifold has subtle global obstructions
for gluing multi-instantons (see Chap. 7 in \cite{book-dokr}).
Let us briefly recapitulate this aspect of the stability.

Consider an Einstein manifold $(M,g)$. The metric on $M$ takes the form
\begin{equation}\label{4-metric}
  ds_4^2 = g_{\mu\nu} (x) dx^\mu dx^\nu = e^a \otimes e^a.
\end{equation}
Using the metric, one can determine the spin connections ${\omega^a}_{b} = {\omega^a}_{b \mu} dx^\mu$
and curvature tensors ${R^a}_{b} = \frac{1}{2} {R^a}_{b \mu \nu} dx^\mu \wedge dx^\nu$
by solving the structure equations \cite{egh-report,nakahara}
\begin{eqnarray} \label{t-free}
  &&  T^a = de^a + {\omega^a}_{b} \wedge e^b =0, \\
  \label{curv-eq}
  && {R^a}_{b} = d {\omega^a}_{b} + {\omega^a}_{c} \wedge {\omega^c}_{b}.
\end{eqnarray}
An underlying idea is that gravity can be formulated as a gauge theory of Lorentz group where
spin connections play a role of gauge fields and Riemann curvature tensors correspond to their field
strengths \cite{utiyama}. Another important point is that Riemann curvature tensors ${R^a}_{b}$ are $spin(4)$-valued
two-forms in $\Omega^2 (M) = \Lambda^2 T^* M$. These facts are combined
with the well-known theorems (see Chap. 13 in \cite{book-besse} and Chaps. 1 \& 2 in \cite{book-dokr}):

{\bf Self-duality}. On an orientable Riemannian four-manifold, the 2-forms decompose into the space of self-dual
and anti-self-dual 2-forms,
\begin{equation}\label{sd-asd}
  \Omega^2 = \Omega_+^2 \oplus \Omega_-^2
\end{equation}
defined by the $\pm 1$ eigenspaces of the Hodge star operator $*:  \Omega^2 \to \Omega^2$.

{\bf Lie group isomorphism}. There is a global isomorphism between the four-dimensional Lorentz group
and classical Lie group, i.e., $SO(4)= SU(2)_+ \otimes SU(2)_-/\mathbb{Z}_2$ or $Spin(4)= SU(2)_+ \otimes SU(2)_-$.
It also leads to the splitting of the Lie algebra
\begin{equation}\label{lie-iso}
  so(4) = su_+(2) \oplus su_-(2).
\end{equation}

A central point is that these two decompositions are deeply related to each other due to the canonical vector space
isomorphism between the Clifford algebra $\mathbb{C}l(4)$ in four-dimensions and the exterior algebra
$\Omega^* M = \bigoplus_{k=0}^4 \Lambda^k T^* M$ over a four-dimensional Riemannian manifold $M$ (see Chap. 2 in \cite{book-clifford}).
For the isomorphism between the vector spaces, the chiral operator $\gamma_5 = - \gamma_1 \gamma_2 \gamma_3 \gamma_4$
in the Clifford algebra corresponds to the Hodge-dual operator $*: \Omega^k \to \Omega^{4-k}$ in the exterior algebra.
Indeed the splitting of vector spaces is induced by the existence of the projection operators
\begin{equation}\label{proj-op2}
  P_\pm = \frac{1}{2} (1 \pm *), \qquad P_\pm = \frac{1}{2} (1 \pm \gamma_5)
\end{equation}
acting on the vector space $\Omega^2$ and the $so(4)$ generators $J_{ab} = \frac{1}{4} [\gamma_a, \gamma_b]$, respectively.
(See Appendix A for the explicit matrix representations of $so(4)$ Lie algebra.)
Therefore, the splitting of the two vector spaces in Eqs. \eq{sd-asd} and \eq{lie-iso} is isomorphic each other.

Thus one can apply these decompositions to spin connections and curvature tensors \cite{oh-yang,joy-jhep,opy-jhep}.
The first decomposition is that the spin connections can be split into a pair of $SU(2)_+$
and $SU(2)_-$ gauge fields according to the Lie algebra splitting \eq{lie-iso}:
\begin{equation}\label{ga-split}
\omega_{ab} = A^{(+)i} \eta^i_{ab}  +  A^{(-)i}\overline{\eta}^i_{ab}
\end{equation}
where $ \eta^i_{ab}$ and $\overline{\eta}^i_{ab}$ are the 't Hooft symbols satisfying the self-duality relation
\begin{equation}\label{eta-duality}
  \eta^i_{ab} = \frac{1}{2} \varepsilon_{abcd} \eta^i_{cd},
  \qquad  \overline{\eta}^i_{ab} = - \frac{1}{2} \varepsilon_{abcd} \overline{\eta}^i_{cd}.
\end{equation}
Note that the index $i=(1,2,3)$ refers to the $su(2)_\pm$ Lie algebra index.
Appendix A contains the explicit representation of $so(4)$ Lie algebra and the 't Hooft symbols.
Accordingly the Riemann curvature tensors are also decomposed into a pair of $SU(2)_+$
and $SU(2)_-$ field strengths:
\begin{equation}\label{gf-split}
R_{ab} = F^{(+)i} \eta^i_{ab}  +  F^{(-)i}\overline{\eta}^i_{ab},
\end{equation}
where $SU(2)_\pm$ field strengths are two-forms on $M$ defined by
\begin{eqnarray}\label{2f-form}
   F^{(\pm)i} &=& \frac{1}{2}  F^{(\pm)i}_{cd} e^c \wedge d^d \xx
   &=& dA^{(\pm)i} - \varepsilon^{ijk} A^{(\pm)j} \wedge A^{(\pm)k}.
\end{eqnarray}

The second decomposition \eq{sd-asd} is that the six-dimensional vector space of two-forms canonically splits into the sum
of three-dimensional vector spaces of self-dual and anti-self-dual two-forms.
Canonical bases of self-dual and anti-self-dual two forms are given by
\begin{equation}\label{2sd-asd}
  \zeta^i_+ = \frac{1}{2} \eta^i_{ab} e^a \wedge e^b, \qquad \zeta^i_- = \frac{1}{2} \bar{\eta}^i_{ab} e^a \wedge e^b.
\end{equation}
Using these bases, one can decompose the $SU(2)_\pm$ field strengths in Eq. \eq{2f-form} as
\begin{equation} \label{dec-fs}
  F^{(+)i} = f^{ij}_{(++)} \zeta^j_+ +  f^{ij}_{(+-)} \zeta^j_-, \qquad
  F^{(-)i} =  f^{ij}_{(-+)} \zeta^j_+ +  f^{ij}_{(--)} \zeta^j_-,
\end{equation}
where the canonical bases in \eq{2sd-asd} satisfy the Hodge-duality equation
\begin{equation} \label{2-hodge}
  * \zeta^i_\pm  = \pm \zeta^i_\pm.
\end{equation}
Combining the two decompositions \eq{gf-split} and \eq{dec-fs} leads to an irreducible decomposition
of general Riemann curvature tensor \cite{s-duality,book-besse,oh-yang,joy-jhep}
\begin{equation}\label{decom-riem}
  R_{abcd} = f^{ij}_{(++)}\eta^i_{ab}\eta^j_{cd} + f^{ij}_{(+-)}\eta^i_{ab} \bar{\eta}^j_{cd}
  +  f^{ij}_{(-+)} \bar{\eta}^i_{ab} \eta^j_{cd} + f^{ij}_{(--)} \bar{\eta}^i_{ab}\bar{\eta}^j_{cd}.
\end{equation}
The torsion-free condition \eq{t-free} leads to an integrability condition, the so-called first Bianchi identity
\begin{equation}\label{1-bianchi}
 R_{abcd}+ R_{acdb} + R_{adbc} = 0.
\end{equation}
From the first Bianchi identity \eq{1-bianchi}, one can derive the symmetry property
\begin{equation}\label{symm-riem}
 R_{abcd} = R_{cdab}.
\end{equation}
Eq. \eq{symm-riem}, being totally 15 conditions, imposes the symmetry property
\begin{equation}\label{symm-fij}
f^{ij}_{(++)} = f^{ji}_{(++)}, \quad f^{ij}_{(--)}=f^{ji}_{(--)},
\quad  f^{ij}_{(+-)} = f^{ji}_{(-+)}.
\end{equation}
The first Bianchi identity \eq{1-bianchi}, being totally 16 conditions, imposes an additional constraint
\begin{equation}\label{1-more}
  f^{ij}_{(++)} \delta^{ij} = f^{ij}_{(--)} \delta^{ij}
\end{equation}
that is equivalently written as
\begin{equation}\label{pseudo-r}
  \varepsilon^{abcd} R_{abcd} = 0.
\end{equation}

If $(M, g)$ is an Einstein manifold satisfying the equations $R_{\mu\nu} = \lambda g_{\mu\nu}$ with $\lambda$
a cosmological constant, one can show (\textbf{6.32} in \cite{book-besse} and \cite{oh-yang,joy-jhep}) that
\begin{equation}\label{e-man}
f^{ij}_{(+-)} = 0 = f^{ij}_{(-+)}.
\end{equation}
In this case, the Riemann curvature tensor \eq{gf-split} is a direct sum of self-dual $SU(2)_+$ field strengths
and anti-self-dual $SU(2)_-$ field strengths taking the form
\begin{equation} \label{em-sd}
  F^{(+)i} = f^{ij}_{(++)} \zeta^j_+ , \qquad
  F^{(-)i} =  f^{ij}_{(--)} \zeta^j_-.
\end{equation}
This means that $SU(2)_\pm$ field strengths describing an Einstein manifold correspond to Yang-Mills instantons
obeying the self-duality equations explicitly written as
\begin{equation}\label{sdasd-eq}
  F^{(\pm) i}_{\mu\nu} = \pm \frac{1}{2} \frac{\varepsilon^{\alpha\beta\rho\sigma}}{\sqrt{g}}
  g_{\mu \alpha} g_{\nu \beta} F^{(\pm) i}_{\rho\sigma}
\end{equation}
where $F^{(\pm) i}= \frac{1}{2} F^{(\pm) i}_{\mu\nu} dx^\mu \wedge dx^\nu
= \frac{1}{2} F^{(\pm) i}_{ab} e^a \wedge e^b$.
Therefore, an Einstein manifold $(M, g)$ has a configuration consisting of $SU(2)_+$ Yang-Mills instantons
and $SU(2)_-$ anti-instantons \cite{oh-yang}.

Since Einstein manifolds encode a topological information in the form of Yang-Mills instantons,
it is natural to expect that the topological invariants of an Einstein manifold $(M,g)$ will be determined
by the configuration of $SU(2)_\pm$ Yang-Mills instantons.
For a general closed Riemannian manifold $M$, the Euler characteristic $\chi(M)$ and
the Hirzebruch signature $\tau(M)$ are defined by \cite{book-besse,egh-report}
\begin{eqnarray} \label{top-euler}
\chi(M) &=& \frac{1}{32 \pi^2} \int_M \varepsilon^{abcd} R_{ab} \wedge R_{cd}, \\
\label{top-hirze}
\tau(M) &=& \frac{1}{24 \pi^2} \int_M  R_{ab} \wedge R_{ab}.
\end{eqnarray}
The topological invariants can be expressed in terms of $SU(2)_\pm$ gauge fields
using the decompositions \eq{gf-split} and \eq{dec-fs}
\begin{eqnarray} \label{dec-euler}
\chi(M) &=& \frac{1}{4 \pi^2} \int_M \big( F^{(+)i} \wedge F^{(+)i} - F^{(-)i} \wedge F^{(-)i} \big), \xx
&=& \frac{1}{2 \pi^2} \int_M \left( (f^{ij}_{(++)})^2 + (f^{ij}_{(--)})^2 - 2 (f^{ij}_{(+-)})^2 \right) \sqrt{g} d^4 x, \\
\label{dec-hirze}
\tau(M) &=& \frac{1}{6 \pi^2} \int_M \big( F^{(+)i} \wedge F^{(+)i} + F^{(+)i} \wedge F^{(+)i} \big), \xx
&=& \frac{1}{3 \pi^2} \int_M \left( (f^{ij}_{(++)})^2 - (f^{ij}_{(--)})^2 \right) \sqrt{g} d^4 x,
\end{eqnarray}
where we used the volume element
\begin{equation*}
  \zeta_\pm^i \wedge \zeta_\pm^j = \pm 2 \delta^{ij} \sqrt{g} d^4 x, \qquad
  \zeta_\pm^i \wedge \zeta_\mp^j = 0.
\end{equation*}
An Einstein manifold has curvature tensors given by \eq{em-sd} with the coefficients satisfying \eq{1-more}.
In this case, the Euler characteristic $\chi(M)$ is given by the sum of self-dual and anti-self-dual instantons
whereas the Hirzebruch signature $\tau(M)$ is their difference. The above expression immediately verifies
the famous inequalities for the topological invariants. The first inequality is $\chi(M)\geq 0$ with equality
only if $f^{ij}_{(++)} = f^{ij}_{(--)} = 0$, i.e., $M$ is flat (\textbf{6.32} in \cite{book-besse} and Sect. 10.4 in \cite{egh-report}).
The second inequality is the Hitchin-Thorpe inequality \cite{hitchin1974} stating that
\begin{equation}\label{hitchin-thorpe}
  \chi (M) \pm \frac{3}{2} \tau (M) = \frac{1}{\pi^2} \int_M  \big(f^{ij}_{(\pm\pm)} \big)^2  \sqrt{g} d^4 x \geq 0
\end{equation}
where the equality holds only if $f^{ij}_{(++)} = 0$ or $f^{ij}_{(--)} = 0$, i.e., $M$ is half-flat
(a gravitational instanton).

\begin{figure}
\centering
\includegraphics[width=0.6\textwidth]{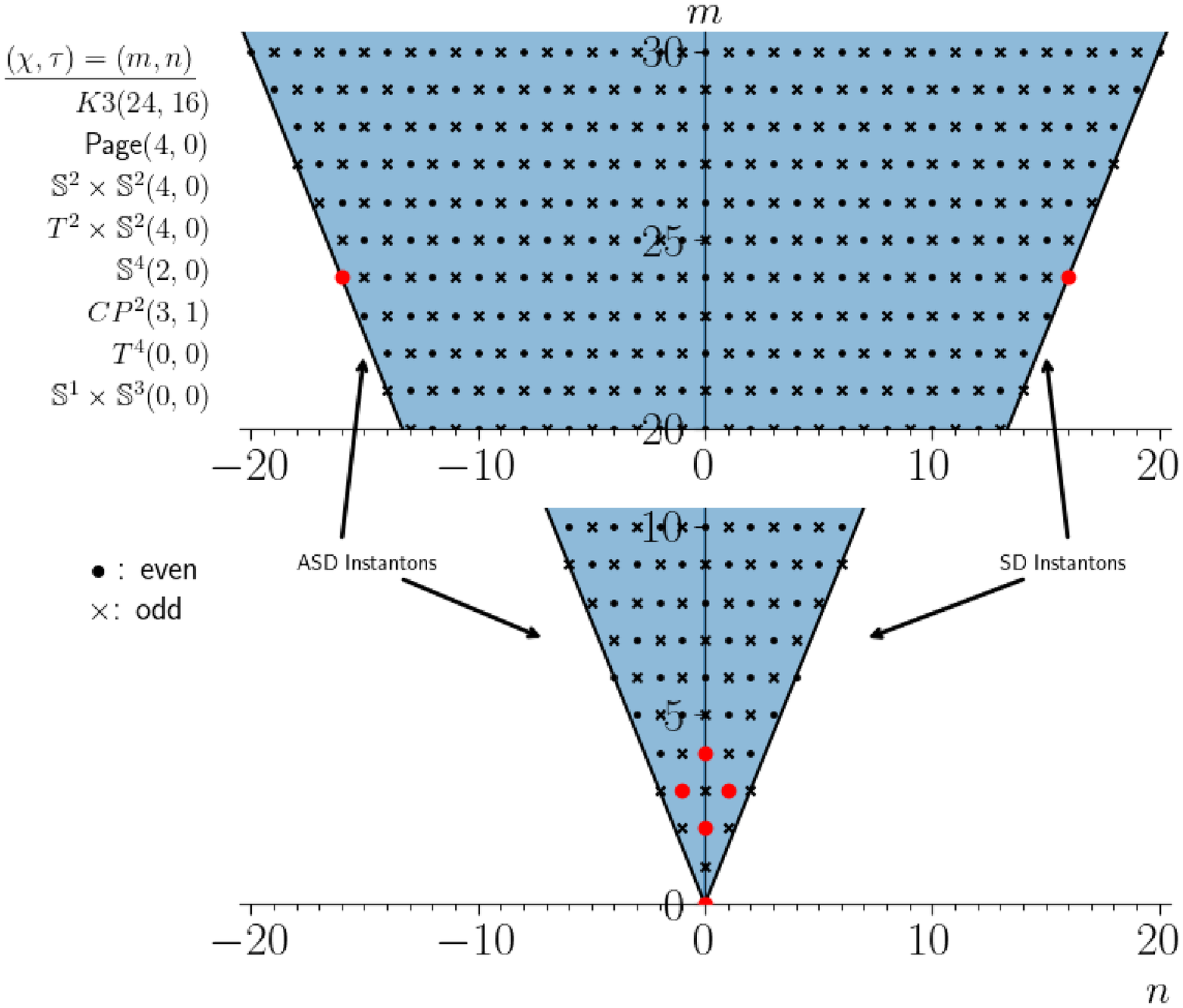}
\caption{Topological numbers of closed Einstein manifolds}
\label{top-numbers}
\end{figure}

The instanton number for $SU(2)_\pm$ gauge fields is defined by\footnote{This definition has considered
the fact \cite{opy-jhep} that $SU(2)_\pm$ gauge fields from spin connections in \eq{ga-split}
are related to Yang-Mills gauge fields by $A^i_G = - \frac{1}{2} A^i_{YM}$
and $F^i_G = - \frac{1}{2} F^i_{YM}$ and $SU(2)$ generators in gravity and gauge theory are related by $T^i_G = - 2 \tau^i_{YM}$.
Note that the $4 \times 4$ matrices $T^i_G = \eta^i_{ab}$ or $\overline{\eta}^i_{ab}$
correspond to the spin $s=\frac{3}{2}$ representation of $SU(2)$ Lie algebra as shown in Eqs. \eq{t+} and \eq{t-}
while the $2 \times 2$ Pauli matrices $\tau^i = 2i \tau^i_{YM}$ in $SU(2)$ gauge fields are the spin $s=\frac{1}{2}$ representation.}
\begin{equation}\label{inst-number}
  I^{(\pm)} = \pm \frac{1}{4 \pi^2}\int_M F^{(\pm)i} \wedge F^{(\pm)i}.
\end{equation}
Then the topological invariants are determined by $SU(2)_\pm$ instantons
\begin{equation} \label{top-inst}
\chi(M) = \big( I^{(+)} + I^{(-)} \big) \geq 0, \qquad
\tau (M) = \frac{2}{3}  \big( I^{(+)} - I^{(-)} \big).
\end{equation}
Let $\chi(M) = m \in \mathbb{Z}_{\geq 0}$ and $\tau(M) = n \in \mathbb{Z}$.
We can invert Eq. \eq{top-inst} as
\begin{equation}\label{inst-mn}
 I^{(\pm)} = \frac{1}{4} \left( 2 m \pm 3 n \right).
\end{equation}
Note that our sign convention in Eq. \eq{inst-number} is $I^{(\pm)} \geq 0$, so the relation \eq{inst-mn}
is consistent with the inequality \eq{hitchin-thorpe}.
The above relations show how the topology of Einstein manifolds is characterized by the configuration
of $SU(2)_+$ instantons and $SU(2)_-$ anti-instantons. One can also deduce that $\chi(M) + \tau (M)
= m + n = 2(1- b_1 + b_2^+) \in 2 \mathbb{Z}$ where $b_i = \mathrm{dim} H^i (M, \mathbb{R})$
is the $i$-th Betti number (Chap. 6.D in \cite{book-besse} and Sect. 10.4 in \cite{egh-report}).
This means that the set $(m,n)$ of topological numbers forms an even integer lattice, i.e.,
$m + n \in 2 \mathbb{Z}$. Some examples of four-dimensional compact Einstein manifolds
are shown up in Fig. \ref{top-numbers} where the structure of the inverted triangle for an allowed region is due to the inequalities
$m \geq 0$ and $m \geq  \frac{3}{2} |n|$. In the list in Fig. \ref{top-numbers},
$\mathbb{S}^1 \times \mathbb{S}^3$ is not an Einstein manifold
since it does not admit Einstein metrics \cite{hitchin1974}
and Page is an inhomogeneous Einstein metric on the product of the nontrivial $\mathbb{S}^2$ bundle
over $\mathbb{S}^2$ \cite{page-metric}.

The Figure \ref{top-numbers} clearly shows the ``reflection" symmetry \cite{pos-yang}.
The ``reflection" symmetry can be realized by considering two compact Einstein manifolds $(M, g)$
and $(\widetilde{M}, \widetilde{g})$ obeying the following relation
\begin{equation}\label{ref-symm}
  I^{(+)} (M)  = I^{(-)} (\widetilde{M}), \qquad  I^{(-)} (M)  = I^{(+)} (\widetilde{M}).
\end{equation}
Under the above transformation \eq{ref-symm}, the topological invariants are related as
\begin{equation}\label{ref-top}
  \chi(M) = \chi(\widetilde{M}), \qquad \tau(M) = - \tau(\widetilde{M}).
\end{equation}
Thus the reflection symmetry corresponds to the interchange of instantons and anti-instantons
which is achieved by a change of the manifold's orientation.
This map indicates that a four-manifold with $\tau(M) = 0$ is self-mirror.

We take the Lie algebra generators of $SO(4)$ as
\begin{eqnarray} \label{so-4t}
J_{ij} \equiv \varepsilon_{ijk} J_k, \qquad J_{i4} \equiv N_{i},
\end{eqnarray}
where $i,j,k =1,2,3$ and
\begin{eqnarray} \label{abcdt}
&& J_{i} = \frac{i}{2} (\tau^i \otimes \mathbf{1}_2) = \frac{i}{2} \left(
                                   \begin{array}{cc}
                                     \tau^i & 0 \\
                                     0 & \tau^i \\
                                   \end{array}
                                 \right), \qquad
N_{i} = \frac{i}{2} (\tau^i \otimes \tau^3) = \frac{i}{2} \left(
                                   \begin{array}{cc}
                                     \tau^i & 0 \\
                                     0 & - \tau^i \\
                                   \end{array}
                                      \right).
\end{eqnarray}
They satisfy the commutation relations
\begin{equation}\label{jncomm}
  [J_i, J_j] = - \varepsilon^{ijk} J_k, \qquad [N_i, N_j] = - \varepsilon^{ijk} J_k,
  \qquad [J_i, N_j] = - \varepsilon^{ijk} N_k.
\end{equation}
In this representation, the generators in the Cartan subalgebra are $-i J_3$ and $-i N_3$.
An irreducible representation (irrep) of $SO(4)$ is labeled by the highest weight defined
by these operators \cite{book-wybo},
which is denoted by a state
\begin{equation}\label{high-w-so4}
  \left|\frac{p}{2}, \frac{q}{2} \right\rangle, \qquad  p \geq |q|
\end{equation}
where $p$ and $q$ are both even integers or both odd integers.
The isomorphism $Spin (4) \cong SU(2)_+ \otimes SU(2)_-$ can be realized by taking
\begin{eqnarray} \label{2su2}
J^{(+)}_i \equiv \frac{1}{2} (J_i + N_i), \qquad J^{(-)}_i \equiv \frac{1}{2} (J_i - N_i),
\end{eqnarray}
because they separately obey the $su(2)\cong so(3)$ commutation relations
\begin{equation}\label{2su2-comm}
  [J^{(\pm)}_i, J^{(\pm)}_j] = - \varepsilon^{ijk} J^{(\pm)}_k, \qquad [J^{(\pm)}_i, J^{(\mp)}_j] = 0.
\end{equation}
For each $SU(2)_\pm$ factor, one may take a spin-$j_\pm$ representation such that $J^{(\pm)}_i J^{(\pm)}_i = - j_\pm (j_\pm +1)$.
We choose $j_+ \geq j_-$.
An $SO(4)$ irrep in this basis is then labeled by a pair of integers or half integers $(j_+, j_-)$, i.e.,
the angular momenta associated with the $su(2)_+$ and $su(2)_-$ subalgebras.
We denote the highest weight state as
\begin{equation}\label{high-w-pq}
  \left|j_+, j_- \right\rangle.
\end{equation}
The two representations are related by putting $p = 2 (j_+ + j_-)$ and $q = 2(j_+ - j_-)$ (Sect. 19.13, \cite{book-wybo}).
The irrep of the direct product $\mathcal{D}(j_+) \otimes \mathcal{D}(j_-)$
is decomposed as
\begin{equation}\label{so-irrep}
  \left[\frac{p}{2}\,, \frac{q}{2} \right] \to \left[\frac{p}{2} \right] \oplus
  \left[\frac{p}{2} - 1 \right] \oplus \cdots \oplus \left[\frac{|q|}{2} \right]
\end{equation}
under the restriction $SO(4) \to SO(3)$.

We note that the separation of instantons such as Eq. \eq{em-sd}
is caused by the splitting of the Lie algebra \eq{lie-iso}.
Considering the fact that the instanton action $S= 8 \pi^2 |I^{(\pm)}|$
is determined by the instanton number itself, it may be reasonable to identity
the instanton numbers $I^{(\pm)}$ with the labels characterizing some irreps of $SU(2)_\pm$.
Now we identify the labels $(j_+, j_-)$ in the representation \eq{high-w-pq}
with the instanton numbers in \eq{inst-mn} as follows:
\begin{equation}\label{id-inst-pq}
 j_+ = 2  I^{(+)} = \frac{1}{2} (2m + 3n) \geq 0, \qquad
 j_- = 2  I^{(-)} = \frac{1}{2} (2m - 3n) \geq 0.
\end{equation}
This identification automatically satisfies the Hitchin-Thorpe inequality \eq{hitchin-thorpe}.
However, in order to satisfy the condition that the set $(m,n)$ of topological numbers forms an even integer lattice,
i.e., $m + n \in 2 \mathbb{Z}$, it is necessary to choose $(j_+, j_-)$ such that
both are integers or half integers and $5j_+ + j_- \in 12 \mathbb{Z}$.
This identification leads to the identification
$p = 2 (j_+ + j_-) = 4m = 4 \chi \geq 0$ and $q = 2 (j_+ - j_-) =6n = 6 \tau$ in the representation \eq{high-w-so4}.
The condition, $m + n \in 2 \mathbb{Z}$, corresponds to the requirement that both $p$ and $q$ are even
and $3p + 2q \in 24 \mathbb{Z}$.
The reflection symmetry \eq{ref-symm} corresponds to the interchange of the representations, $(j_+ \leftrightarrow j_-)$,
under which $(p, q) \to (p, - q)$. This is the reason why it is enough to consider only the case, $j_+ \geq j_-$,
i.e. $q \geq 0$. Note that the representations $\mathcal{D}(j_+) \otimes \mathcal{D}(j_-)$ and
$\mathcal{D}(j_-) \otimes \mathcal{D}(j_+)$ for $j_+ \neq j_-$ correspond to distinct representations in $SO(4)$.
It may be pointed out that the identification \eq{id-inst-pq} does not explain the Hitchin-Thorpe inequality \eq{hitchin-thorpe}
because we have chosen the representations $(j_+, j_-)$ such that they obey the relation $ j_+ = 2  I^{(+)}$ and
$j_- = 2  I^{(-)}$. Nevertheless, it is very encouraging that it is always possible to choose
the $SO(4)$ representations so that such a relation is satisfied.

The four-dimensional Lorentz group $Spin(4)$ is the spin group in dimension 4, the double cover of $SO(4)$,
that is a product group since $Spin(4) = SU(2)_+ \times SU(2)_-$
and its Lie algebra becomes a direct sum of two $su(2)_\pm$ Lie algebras.
The splitting of the Lie algebra in Eq. \eq{lie-iso} is related to the decomposition
of the 2-forms on a four-manifold in Eq. \eq{sd-asd}.
The canonical splitting of the vector spaces occupies a central position for the instanton structure of Einstein manifolds.
However, one may think of the four-dimensional gravity as being obtained through the Kaluza-Klein reduction
of a five-dimensional gravity. Then one can consider the four-dimensional Einstein manifolds as obtained
from a five-dimensional gravitational solution.
The five-dimensional Lorentz group is $SO(5)$ which is a simple group.
Since $SO(4) \subset SO(5)$, it will be interesting to see how the instanton configuration of Einstein manifolds
fits into a multiplet in the irrep of $SO(5)$.
We will discuss this issue in the next section.

\section{Representation of Riemannian manifolds}

As we have observed in Sec. 2, the separation of Riemann curvature tensors
has been originated from the splitting of the Lie algebra vector space \eq{lie-iso}.
Since the vector space $so(4)$ is isomorphic to the vector space of two-forms $\Omega^2$,
the same kind of splitting must also arise in the vector space $\Omega^2$.
Eq. \eq{sd-asd} precisely shows such splitting of the vector space $\Omega^2$.
The gravitational force is represented by Riemann curvature tensors, and Einstein manifolds
are described by two independent components of Riemann tensors (i.e., self-dual and anti-self-dual
gravitational instantons).\footnote{It is not the case for the Lorentzian signature
because the local Lorentz group $SO(3,1)$ is a simple group although it is not compact.
Furthermore, it does not mean that there are two independent gravitational forces because the gravitational force
is transmitted by metrics (not connections) and the metric does not decompose in any sense into a sum of two independent parts.
The self-dual and anti-self-dual components of spin connections are described by the same metric.}
However, an interesting physics arises if we consider the four-dimensional gravity as being obtained
from a five-dimensional gravity through the Kaluza-Klein compactification \cite{kk-book}.
Then the five-dimensional Lorentz group is $SO(5)$ that is a simple group unlike the group
$SO(4)= SU(2)_+ \times SU(2)_-/\mathbb{Z}_2$. Moreover, there is no concept of self-duality for two-forms
in five dimensions so that the vector space $\Omega^2$ is no more decomposed.
Therefore, neither the Lie algebra of $SO(5)$ nor the vector space of two-forms $\Omega^2$ is splitted in five dimensions.
This implies that the self-dual and anti-self-dual components in $SU(2)_\pm$ factors must be combined in five dimensions
since the group $SO(4) = SU(2)_+ \times SU(2)_-/\mathbb{Z}_2$
has to be embedded into the simple group $SO(5)$.
In other words, $SU(2)_+$ instantons and $SU(2)_-$ anti-instantons in four-dimensional Einstein manifolds
must appear in the same multiplet of the Lorentz group $SO(5)$.
Therefore the five-dimensional Kaluza-Klein theory unifies two independent sectors of curvature tensors
as well as the electromagnetic force into a single gravitational force.

This scheme is similar to the grand unification of Standard Model although the Kaluza-Klein theory
is defined in five-dimensional space.
The Standard Model has a product gauge group $SU(3) \times SU(2) \times U(1)$ to describe the electroweak and strong forces.
In the GUT, the Standard Model gauge group is embedded into a single gauge group, for example,
$SU(5)$ or $SO(10)$. Then the leptons and quarks appear in the same multiplet in a larger symmetry \cite{georgi-book}.
Now we will see how four-dimensional Riemannian manifolds are similarly combined into a five-dimensional
Einstein manifold. Furthermore, we will see that $SU(2)_+$ instantons and $SU(2)_-$ anti-instantons play a role of
quarks and anti-quarks from the point of view of a five-dimensional Einstein manifold.
In order to analyze the anatomy of Riemannian manifolds, we will greatly use the group isomorphism \cite{penn4}
\begin{equation}\label{2-group-iso}
  SO(5) \cong Sp(2)/\mathbb{Z}_2.
\end{equation}
We provide more details about the Lie algebras $so(4)=su(2)_+ \oplus su(2)_-$ and $so(5)\cong sp(2)$ in Appendix A.

Let $N$ be a five-dimensional Riemannian manifold whose metric is given by
\begin{equation}\label{5-metric}
  ds^2 = G_{MN} (X) dX^M dX^N.
\end{equation}
Introduce at each spacetime point on $N$ a local basis of orthonormal tangent vectors
$\mathcal{E}_A = \mathcal{E}_A^M \partial_M \in \Gamma(TN)$ and its dual basis $E^A = E^A_M dX^M \in \Gamma(T^*N)$
defined by a natural pairing $\langle E^A, \mathcal{E}_B \rangle = \delta^A_B$ where $A, B = 1, \cdots, 5; M, N = 1, \cdots, 5$.
In terms of the non-coordinate basis in $\Gamma(T^*N)$, the metric \eq{5-metric} can be written as
\begin{equation}\label{5-metric-viel}
  ds^2 = G_{MN} (X) dX^M dX^N = \delta_{AB} E^A \otimes E^B.
\end{equation}
Let us consider Einstein manifolds $(N, G)$ described by the Einstein-Hilbert action
\begin{equation}\label{5-ehaction}
  S_5 = - \frac{1}{16 \pi G_5} \int (R - 3 \Lambda) \sqrt{G} d^5 X
\end{equation}
where $\Lambda$ and $G_5$ are a cosmological constant and the gravitational constant
in five dimensions, respectively. The equations of motion derived from the action \eq{5-ehaction} are
\begin{equation}\label{5ein-eq}
  R_{MN} = \Lambda G_{MN}.
\end{equation}
A solution to Eq. \eq{5ein-eq} constitutes five-dimensional Einstein manifolds $(N, G)$.
Now we consider the Kaluza-Klein compactification of five-dimensional Einstein manifolds by assuming that
the five-dimensional space $N$ is a cylinder $M \times \mathbb{S}^1$ with $0 \leq x^5 \leq L = 2\pi R_5$ \cite{kk-book}.
We split five-dimensional coordinates as $X^M = (x^\mu, x^5), \; \mu=1, \cdots, 4$, according to the cylinder geometry.
Then the five-dimensional metric tensor in \eq{5-metric} also splits into four-dimensional fields, $g_{\mu\nu} (x), \; A_\mu (x)$
and $\phi(x)$. We have imposed the “cylinder condition” that the fields should not depend on the fifth coordinate $x^5$.
We take the Kaluza-Klein ansatz for the five-dimensional metric in the form
\begin{eqnarray}\label{kk-metric}
  ds^2 &=& G_{MN} (X) dX^M dX^N \xx
  &=& e^{-\frac{1}{3} \phi} \left( g_{\mu\nu} dx^\mu dx^\nu + e^\phi (dx^5 + \kappa A_\mu dx^\mu)^2 \right)
\end{eqnarray}
where $\kappa^2 = 16 \pi G_4$ and $G_4 = \frac{G_5}{L}$ is the four-dimensional gravitational constant.
It may be instructive to write the five-dimensional metric in the matrix form
\begin{equation}\label{5-matrix}
G_{MN} = e^{-\frac{1}{3} \phi} \left(
           \begin{array}{cc}
             g_{\mu\nu} + \kappa^2  e^\phi A_\mu A_\nu  &  \kappa e^\phi A_\mu \\
             \kappa  e^\phi A_\nu & e^\phi \\
           \end{array}
         \right).
\end{equation}
The geometric details of the five-dimensional gravity and the Kaluza-Klein theory appear in Appendix B.

Using the result \eq{kk-54}, one can write down the five-dimensional Einstein-Hilbert action \eq{5-ehaction}
for the Kaluza-Klein ansatz \eq{kk-metric}.
First note that the five-dimensional volume form is $\sqrt{G} d^5 X = E^1 \wedge \cdots \wedge E^5
= e^{-\frac{1}{3}\phi} \sqrt{g} d^4 x dx^5$. Since the four-dimensional fields do not depend on the circle coordinate $x^5$,
one can integrate out the fifth coordinate that gives rise to the redefinition of the gravitational constant $G_4 = \frac{G_5}{L}$.
Moreover, one can ignore the Laplacian term in \eq{kk-54} because it becomes a boundary term.
Finally, the Einstein-Hilbert action \eq{5-ehaction} reduces to the four-dimensional action
\begin{equation}\label{kk-ehaction}
  S = \int \Big( - \frac{1}{16 \pi G_4} \big( \leftidx{^{(4)}}{R} - 3 e^{- \frac{\kappa}{\sqrt{3}} \Phi} \Lambda \big)
  + \frac{1}{4} e^{\sqrt{3} \kappa \Phi} g^{\mu\rho} g^{\nu\sigma} F_{\mu\nu} F_{\rho\sigma}
  + \frac{1}{2} g^{\mu\nu} \partial_\mu \Phi \partial_\nu \Phi \Big) \sqrt{g} d^4 x,
\end{equation}
where we have rescaled the scalar field
\begin{equation}\label{res-phi}
  \Phi \equiv \frac{1}{\sqrt{3} \kappa} \phi
\end{equation}
such that the scalar field has the usual kinetic term with canonical mass dimension.
The Ricci scalar $\leftidx{^{(4)}}{R}$ with the left-hand superscript $\leftidx{^{(4)}} \,$ is determined
only by the four-dimensional metric $ds_4^2 = g_{\mu\nu} (x) dx^\mu dx^\nu$.
Note that $\Lambda$ no longer behaves like a cosmological constant in four dimensions
except the case of a constant scalar field.

Let us consider the symmetries of the Kaluza-Klein geometry with the metric \eq{kk-metric} where the components of
the gravitational field along the circle transmute into the electromagnetic field.
The effective field theory of five-dimensional gravity around a solution of the form $N = M \times \mathbb{S}^1$
is four-dimensional gravity coupled to electromagnetism and a dilaton field.
The five-dimensional Lorentz transformations that would mix the remaining four-dimensional gravitational excitations
with electromagnetic excitations are not symmetries of the metric.
The symmetries of the Kaluza-Klein vacuum \eq{kk-metric} are the four-dimensional Lorentz symmetries, acting on $M$,
and a $U(1)$ group acting on the circle $\mathbb{S}^1$  \cite{kk-witten,kk-book}.
These symmetries are realized as local or gauge symmetries in the apparent four-dimensional world
because the whole theory started with the Einstein-Hilbert action \eq{5-ehaction}
which is generally covariant. Therefore the spontaneous symmetry breaking by the Kaluza-Klein ground state \eq{kk-metric}
arises via a two step procedure with the symmetry breaking from $SO(5)$ to $SO(4)$ followed by the symmetry enhancement to
$SO(4) \times U(1)$ in terms of the isometry of the Kaluza-Klein circle.
The remaining symmetry is denoted as
\begin{equation}\label{symm-break}
  SO(5) \to SO(4) \times U(1),
\end{equation}
although the $U(1)$ factor is not a subgroup of $SO(5)$ since it acts on the circle coordinate as
\begin{equation}\label{5th-u1}
  x^\mu \mapsto x^\mu, \qquad x^5 \mapsto x^5 + f(x).
\end{equation}
Under this transformation, we have
\begin{equation}\label{kk-tran}
  g_{\mu\nu} \mapsto  g_{\mu\nu}, \qquad \phi \mapsto \phi,
  \qquad A_{\mu} \mapsto  A_{\mu} - \frac{1}{\kappa} \partial_\mu f,
\end{equation}
so that the one-form $A = A_\mu dx^\mu$ transforms like an Abelian gauge field.

The equations of motion for the four-dimensional fields can be derived from the action \eq{kk-ehaction}:
\begin{eqnarray} \label{4eom-e}
&& \leftidx{^{(4)}}{R}_{\mu\nu} - \frac{1}{2} g_{\mu\nu} \big( \leftidx{^{(4)}}{R}
- 3 e^{- \frac{\kappa}{\sqrt{3}} \Phi} \Lambda \big)
= 8\pi G_4 T_{\mu\nu}, \\
\label{4eom-m}
&&  D_\mu \left( e^{\sqrt{3} \kappa \Phi} F^{\mu\nu} \right) = 0, \\
\label{4eom-s}
&&  \Delta \Phi = \frac{\sqrt{3}\kappa}{4} e^{\sqrt{3} \kappa \Phi} F_{\mu\nu} F^{\mu\nu}
- \frac{\sqrt{3}}{\kappa} e^{- \frac{\kappa}{\sqrt{3}} \Phi} \Lambda,
\end{eqnarray}
where the energy-momentum tensor $T_{\mu\nu}$ is given by
\begin{equation}\label{em-tensor}
 T_{\mu\nu} = e^{\sqrt{3} \kappa \Phi} g^{\rho\sigma}  F_{\mu\rho} F_{\nu\sigma}
 + \partial_\mu \Phi \partial_\nu \Phi
 - g_{\mu\nu} \Big( \frac{1}{4} e^{\sqrt{3} \kappa \Phi} F_{\rho\sigma} F^{\rho\sigma}
  + \frac{1}{2} g^{\rho\sigma} \partial_\rho \Phi \partial_\sigma \Phi \Big).
\end{equation}
Indeed one can check using the results in Eqs. \eq{kk-51}-\eq{kk-54} and $R = 5\Lambda$ that
the above equations of motion are exactly the same as Eq. \eq{5ein-eq} for a five-dimensional Einstein manifold.
Therefore, the general five-dimensional metric \eq{kk-metric} describes a five-dimensional Einstein manifold
as long as the four-dimensional fields, $g_{\mu\nu} (x), \; A_\mu (x)$ and $\Phi(x)$,
satisfy the above equations of motion. It will be interesting to see how the other fields such as
$A_\mu (x)$ and $\Phi(x)$ deform the instanton structure of four-dimensional Einstein manifolds and
understand how these deformed geometries are nicely unified into a five-dimensional Einstein manifold.

In order to understand such a structure of a five-dimensional Einstein manifold, it would be useful
to have the decomposition of five-dimensional Riemann curvature tensors
similar to the four-dimensional decomposition \eq{decom-riem}.
The generators of $so(5)$ Lorentz algebra are defined by
\begin{equation}\label{t5-gen}
  J_{AB} = \frac{1}{4} [\gamma_A, \gamma_B]
\end{equation}
and they satisfy the Lorentz algebra
\begin{equation}\label{t5-dirac}
  [J_{AB}, J_{CD}] = - ( \delta_{AC} J_{BD} - \delta_{AD} J_{BC} - \delta_{BC} J_{AD} + \delta_{BD} J_{AC}).
\end{equation}
See Appendix A for the representation of the five-dimensional gamma matrices.
The 10 generators in Eq. \eq{t5-gen} consist of $J_{AB}=(J_{ab}, J_{5a})$ where
the generators $J_{ab}$ satisfy the four-dimensional Lorentz algebra $so(4) \subset so(5)$ and
$J_{5a}$ are additional generators given by
\begin{eqnarray} \label{tso5-dec}
 J_{ab}  &=&  \frac{1}{4} [\gamma_a, \gamma_b] = \frac{i}{2} \left(
                                                               \begin{array}{cc}
                                                                 \eta^i_{ab} \tau^i  & 0 \\
                                                                 0 & \bar{\eta}^i_{ab} \tau^i \\
                                                               \end{array}
                                                             \right), \xx
 J_{5a}  &=&  \frac{1}{4} [\gamma_5, \gamma_a] = \frac{1}{2} \left(
                                                               \begin{array}{cc}
                                                                 0  &  \sigma^a \\
                                                                 - \bar{\sigma}^a & 0 \\
                                                               \end{array}
                                                             \right).
\end{eqnarray}
Let us denote the generators in Eq. \eq{tso5-dec} as
\begin{eqnarray} \label{tab-45}
  J_{ij} &\equiv& \varepsilon_{ijk} T^k, \qquad \quad \; J_{i4} \equiv T^{3+i},  \xx
  J_{i5} &\equiv& - T^{6+i}, \qquad \quad \; J_{45} \equiv -  T^{10},
\end{eqnarray}
which take the block matrix form:
\begin{eqnarray} \label{tabcd}
&& T^{i} = \frac{i}{2} (\tau^i \otimes \mathbf{1}_2) = \frac{i}{2} \left(
                                   \begin{array}{cc}
                                     \tau^i & 0 \\
                                     0 & \tau^i \\
                                   \end{array}
                                 \right), \qquad
T^{3+i} = \frac{i}{2} (\tau^i \otimes \tau^3) = \frac{i}{2} \left(
                                   \begin{array}{cc}
                                     \tau^i & 0 \\
                                     0 & - \tau^i \\
                                   \end{array}
                                 \right), \xx
 && T^{6+i} = \frac{i}{2} (\tau^i \otimes \tau^1) = \frac{i}{2} \left(
                                   \begin{array}{cc}
                                     0 & \tau^i \\
                                     \tau^i & 0 \\
                                   \end{array}
                                 \right), \qquad
T^{10} = \frac{i}{2} (\mathbf{1}_2 \otimes \tau^2) = \frac{1}{2} \left(
                                   \begin{array}{cc}
                                     0 & \mathbf{1}_2 \\
                                     - \mathbf{1}_2 & 0 \\
                                   \end{array}
                                 \right).
\end{eqnarray}
It can be shown (see Appendix A) that the $4 \times 4$ matrices $T^\mathbb{A}, \; \mathbb{A} = 1, \cdots, 10$, in \eq{tabcd}
constitute the Lie algebra generators of $sp(2)$. Therefore we establish the Lie algebra isomorphism $sp(2) \cong so(5)$.
Since the universal covering group of $SO(5)$ is $Sp(2)$, we get the group isomorphism \eq{2-group-iso}.

Since the vector spaces generated by $J_{AB}$ and $T^\mathbb{A}$ are isomorphic each other,
there exists a linear relation between them:
\begin{equation}\label{linear-rel}
  J_{AB} = \psi^\mathbb{A}_{AB} T^\mathbb{A},  \qquad T^\mathbb{A} = \frac{1}{2} \psi^\mathbb{A}_{AB} J_{AB},
\end{equation}
where
\begin{equation}\label{eta-symbol}
 \psi^\mathbb{A}_{AB} = - Tr (  J_{AB} T^\mathbb{A} ).
\end{equation}
The psi-symbol in \eq{eta-symbol} is the analogue of the four-dimensional 't Hooft symbols in \eq{thooft-matrix}
which explicitly presents the Lie algebra isomorphism $so(5) \cong sp(2)$.
Indeed the matrix expression $(T^\mathbb{A})_{AB} \equiv \psi^\mathbb{A}_{AB}$ provides
the five-dimensional representation of $sp(2)$ Lie algebra as was shown in \eq{5lie-sp2}.
The Riemann curvature tensor $R = \frac{1}{2}  R_{AB} J_{AB} \in C^\infty ( \mathfrak{g} \otimes \Omega^2)$
carries two kinds of indices living in different vector spaces:
\begin{equation}\label{riemann-curvature}
  R_{AB} = \frac{1}{2} R_{ABMN} dX^M \wedge dX^N = \frac{1}{2} R_{ABCD} E^C \wedge E^D
\end{equation}
where the indices $(A,B)$ live in the vector space of $\mathfrak{g}=so(5)$ Lie algebra
while $(C,D)$ live in the vector space of two-forms $\Omega^2 = \Lambda^2 T^* N$.
But these two vector spaces are isomorphic each other and
their isomorphism is encoded in the symmetry property of curvature tensors,
\begin{equation}\label{symm-curv}
  R_{ABCD} = R_{CDAB}.
\end{equation}
The symmetry property \eq{symm-curv} can be derived from the first Bianchi identity
\begin{equation}\label{1st-bianchi}
   R_{ABCD} + R_{ACDB} + R_{ADBC} = 0
\end{equation}
which is the integrability condition for the torsion two-forms in \eq{steq-torsion} \cite{egh-report,nakahara}.
Thus we can expand the curvature tensors $R_{ABCD}$ in the $sp(2)$ basis for both indices
using the psi-symbol \eq{eta-symbol} as
\begin{equation}\label{curv-eta}
  R_{ABCD} = \mathfrak{R}_{\mathbb{A}\mathbb{B}} \psi^\mathbb{A}_{AB} \psi^\mathbb{B}_{CD}
\end{equation}
where the expansion coefficients are symmetric, i.e.,
\begin{equation}\label{r-symm}
\mathfrak{R}_{\mathbb{A}\mathbb{B}}
= \mathfrak{R}_{\mathbb{B}\mathbb{A}} = \frac{1}{4}  R_{ABCD} \psi^\mathbb{A}_{AB} \psi^\mathbb{B}_{CD}
\end{equation}
due to the property \eq{symm-curv}. The Bianchi identity \eq{1st-bianchi} which is totally 50 conditions imposes
five additional conditions
\begin{equation}\label{5-bianchi}
d^{\mathbb{A}\mathbb{B} C}  \mathfrak{R}_{\mathbb{A}\mathbb{B}} = 0
\end{equation}
in addition to the 45 conditions from Eq. \eq{r-symm}, where the structure constants $d^{\mathbb{A}\mathbb{B} C}$
are defined in \eq{anti-sp5}. The constraints \eq{5-bianchi} can be derived
from Eq. \eq{curv-eta} by contracting $\frac{1}{4} J_{AB} J_{CD}$ on both sides and
applying the products \eq{prod-gen1} and \eq{prod-gen2}. Then it results in two identities
\begin{eqnarray} \label{iden-1}
  && \mathfrak{R}_{\mathbb{A}\mathbb{B}} \delta^{\mathbb{A}\mathbb{B}} = \frac{1}{2} R, \\
  \label{iden-2}
   && d^{\mathbb{A}\mathbb{B} E}  \mathfrak{R}_{\mathbb{A}\mathbb{B}} = \frac{1}{4} \varepsilon_{ABCDE} R_{ABCD} = 0,
\end{eqnarray}
where $R$ is the Ricci scalar and Eq. \eq{iden-2} must vanish due to the Bianchi identity \eq{1st-bianchi}.
It may also be checked by counting the number of independent Riemann curvature tensors.
In five dimensions, the number of independent Riemann curvature tensors obeying the Bianchi identity \eq{1st-bianchi} is 50.
The number of $sp(2)$ curvature tensors obeying Eq. \eq{r-symm} is $55=100-45$ and then imposing the five constraints \eq{5-bianchi}
leads to 50 independent components.
See Appendix C for the group structure of Riemann curvature tensor.

\begin{figure}
\centering
\includegraphics[width=0.6\textwidth]{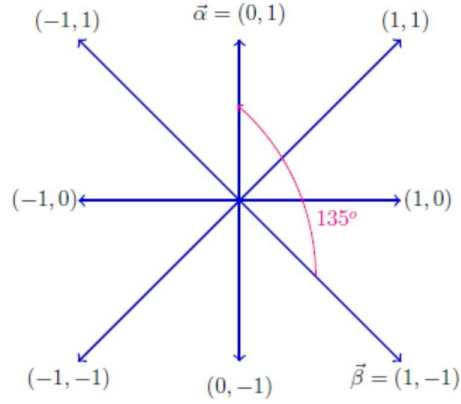}
\caption{Root diagram of $so(5) \cong sp(2)$ for the Cartan subalgebra \eq{cartan}.}
\label{sp2-root1}
\end{figure}

The $sp(2) \cong so(5)$ generators in \eq{tabcd} satisfy the commutation relations
\begin{eqnarray}\label{sp2-comm}
  && [T^{i}, T^{j}] = [T^{3+i}, T^{3+j}] = [T^{6+i}, T^{6+j}] = - \varepsilon^{ijk} T^{k}, \xx
  && [T^{i}, T^{3+j}] = - \varepsilon^{ijk} T^{3+k}, \quad [T^{i}, T^{6+j}] = - \varepsilon^{ijk} T^{6+k},
   \quad [T^{3+i}, T^{6+j}] = - \delta^{ij} T^{10}, \\
  && [T^{i}, T^{10}] = 0, \quad [T^{3+i}, T^{10}] = T^{6+i}, \quad [T^{6+i}, T^{10}] = - T^{3+i}. \nonumber
\end{eqnarray}
Note that $T^3 = \frac{i}{2}(\tau^3 \otimes \mathbf{1}_2)$ and $T^6 = \frac{i}{2}(\tau^3 \otimes \tau^3)$ are diagonal matrices.
Therefore they constitute the set of the Cartan subalgebra for $sp(2) \cong so(5)$
\begin{equation}\label{cartan}
  \mathfrak{h} = \{ H^1 = - i T^3, \; H^2 = - iT^6 \}.
\end{equation}
They correspond to $H^1 = - i J_3$ and $H^2 = - iN_3$, respectively, according to the notation \eq{so-4t}.
Thus this representation contains the highest weight state \eq{high-w-so4}.
The remaining generators are chosen to satisfy the eigenvalue equations \cite{georgi-book,book-wybo}
\begin{equation}\label{root-eq}
  [ H^i, E^\alpha ] = \alpha^i E^\alpha
\end{equation}
where $i=1,2$. The two-dimensional vector $\vec{\alpha} = (\alpha^1, \alpha^2)$ is called a root
and $E^\alpha$ is the corresponding ladder operator.
We choose the ladder operators as follows:
\begin{eqnarray}\label{ladder}
  && A_\pm = \frac{1}{2} \big( T^1 \pm i T^2 + (T^4 \pm i T^5)\big), \quad
  B_\pm = \frac{1}{2} \big( T^1 \pm i T^2 - (T^4 \pm i T^5) \big), \xx
  && C_\pm = T^7 \pm i T^8, \quad
  D_\pm = T^9 \pm i T^{10}.
\end{eqnarray}
It may be useful to show the explicit matrix representation of the Cartan-Weyl basis
\begin{eqnarray} \label{cw-basis}
&& H^{1} = \frac{1}{2} \left(
                                   \begin{array}{cc}
                                     \tau^3 & 0 \\
                                     0 & \tau^3 \\
                                   \end{array}
                                 \right), \quad
 H^{2} = \frac{1}{2} \left(
                                   \begin{array}{cc}
                                     \tau^3 & 0 \\
                                     0 & - \tau^3 \\
                                   \end{array}
                                 \right), \quad
A_\pm =  \frac{1}{2} \left(
                                   \begin{array}{cc}
                                     \tau^\pm & 0 \\
                                     0 & 0 \\
                                   \end{array}
                                 \right),
                                 \quad
B_\pm =  \frac{1}{2} \left(
                                   \begin{array}{cc}
                                     0 & 0 \\
                                     0 & \tau^\pm \\
                                   \end{array}
                                 \right), \xx
&& C_\pm = \frac{1}{2} \left(
                                   \begin{array}{cc}
                                     0 & \tau^\pm \\
                                     \tau^\pm & 0 \\
                                   \end{array}
                                 \right), \qquad
D_\pm = \frac{i}{2} \left(
                                   \begin{array}{cc}
                                     0 & \pm \mathbf{1}_2 + \tau^3 \\
                                     \mp \mathbf{1}_2 + \tau^3 & 0 \\
                                   \end{array}
                                 \right),
\end{eqnarray}
where $ \tau^\pm = i( \tau^1 \pm i \tau^2)$.
Then the commutation relations in Eq. \eq{sp2-comm} can be written in the Cartan-Weyl basis as
\begin{eqnarray}\label{cartan-weyl}
  && [H^1, A_\pm ] = \pm  A_\pm, \quad [H^2, A_\pm ] = 0, \xx
  && [H^1, B_\pm ] = 0, \hspace{1.1cm} [H^2, B_\pm ] = \pm B_\pm, \\
  && [H^1, C_\pm ] = \pm C_\pm, \quad [H^2, C_\pm ] = \pm C_\pm, \xx
  && [H^1, D_\pm ] = \pm D_\pm,  \quad [H^2, D_\pm ] = \mp D_\pm.  \nonumber
\end{eqnarray}
Therefore we identify the root vectors derived from the ladder generators
\begin{equation}\label{rootvector}
  \vec{\alpha}_{A_\pm} = \pm (1,0), \quad \vec{\alpha}_{B_\pm} = \pm (0,1),
  \quad \vec{\alpha}_{C_\pm} = \pm (1,1), \quad
  \vec{\alpha}_{D_\pm} = \pm (1,-1),
\end{equation}
where the first (second) entry of root vectors  is the eigenvalue of $ad_{H^1} \; (ad_{H^2})$.
The corresponding root diagram is shown in Fig. \ref{sp2-root1}. The simple roots are denoted by
\begin{equation}\label{sroot}
  \vec{\alpha}=(0,1), \quad \vec{\beta}=(1,-1).
\end{equation}

\begin{figure}
\centering
\includegraphics[width=0.7\textwidth]{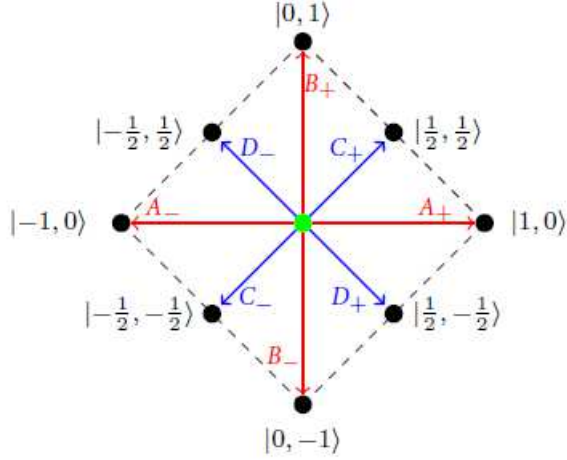}
\caption{The root diagram of $sp(2) \cong so(5)$ for the Cartan subalgebra \eq{cartan-b}
where the actual roots must read as $|\alpha_1, \alpha_2 \rangle
= \sqrt{2}|\beta_1, \beta_2 \rangle$.}
\label{so5-root}
\end{figure}

One may choose a different combination of the Cartan subalgebra
\begin{equation}\label{cartan-b}
  \mathfrak{h} = \{ H^1 = -\frac{i}{\sqrt{2}}(T^3 + T^6), \; H^2 = -\frac{i}{\sqrt{2}}(T^3 - T^6) \}
\end{equation}
whose matrix form is given by
\begin{equation} \label{bcw-basis}
H^{1} = \frac{1}{\sqrt{2}} \left(
                                   \begin{array}{cc}
                                     \tau^3 & 0 \\
                                     0 & 0 \\
                                   \end{array}
                                 \right), \qquad
 H^{2} = \frac{1}{\sqrt{2}} \left(
                                   \begin{array}{cc}
                                     0 & 0 \\
                                     0 & \tau^3 \\
                                   \end{array}
                                 \right).
\end{equation}
They correspond to $H^1 = - \sqrt{2} i J^{(+)}_3$ and $H^2 = - \sqrt{2} i J^{(-)}_3$, respectively,
according to the notation \eq{2su2}.
Thus this representation contains the highest weight state \eq{high-w-pq}.
Then the commutation relations in Eq. \eq{sp2-comm} can be written in the Cartan-Weyl basis as
\begin{eqnarray}\label{cartan-weyl}
  && [H^1, A_\pm ] = \pm \sqrt{2} A_\pm, \quad [H^2, A_\pm ] = 0, \xx
  && [H^1, B_\pm ] = 0, \hspace{1.7cm} [H^2, B_\pm ] = \pm \sqrt{2} B_\pm, \\
  && [H^1, C_\pm ] = \pm \frac{1}{\sqrt{2}} C_\pm, \quad [H^2, C_\pm ] = \pm \frac{1}{\sqrt{2}} C_\pm, \xx
  && [H^1, D_\pm ] = \pm \frac{1}{\sqrt{2}} D_\pm,  \quad [H^2, D_\pm ] = \mp \frac{1}{\sqrt{2}} D_\pm.  \nonumber
\end{eqnarray}
The corresponding root diagram is shown in Fig. \ref{so5-root}. The simple roots may be chosen as
\begin{equation}\label{sroot}
  \vec{\alpha}= \sqrt{2}(0,1), \quad \vec{\beta}=\frac{1}{\sqrt{2}}(1,-1).
\end{equation}
Our normalization for simple roots is that the square length of the longest roots is set equal to 2.

As we have indicated in Eq. \eq{riemann-curvature}, the Riemann curvature tensors are two-forms
in $\Omega^2 = \Lambda^2 T^*N$ taking values in the vector space of $\mathfrak{g} = so(5) \cong sp(2)$ Lie algebra.
According to the remaining symmetry \eq{symm-break},
let us decompose the Lie algebra $\mathfrak{g} = so(5) = \{ T^\mathbb{A}| \mathbb{A}= 1, \cdots, 10 \}$ as
\begin{equation}\label{decom-so5}
  \mathfrak{g} = so(4) \oplus \mathfrak{k} = su(2)_+ \oplus su(2)_- \oplus \mathfrak{k}
\end{equation}
where $\mathfrak{k}$ contains the generators $J_{5a}$ in the coset space $SO(5)/SO(4) \cong \mathbb{S}^4$.
Since $\mathfrak{g} \cong \Lambda^2 T^*N$ as vector spaces, Eq. \eq{curv-eta} gives us the expansion
of the curvature tensor in the basis of $\mathfrak{g}$.
In the basis \eq{tab-45}, $\{T^\mathbb{A}| \mathbb{A}= 1, \cdots, 6 \}$ corresponds to $J_{ab}$,
so $so(4) \subset so(5)$ Lie algebra and $\{T^{6+a}: a = 1, \cdots, 4 \}$ corresponds
to the coset generators $J_{5a}$ in $\mathfrak{k}$.
As is well-known, Eq. \eq{sp2-comm} shows that the coset space $SO(5)/SO(4) \cong \mathbb{S}^4$ is
reductive $([so(4), \mathfrak{k}] \subset \mathfrak{k})$ and symmetric $([\mathfrak{k}, \mathfrak{k}] \subset so(4))$.
One can explicitly determine nonzero components of the psi-symbols
defined by Eq. \eq{eta-symbol} using Eq. \eq{prod-gen1}:
\begin{equation} \label{psi-comp}
  \psi^i_{ab} = \varepsilon^{iab4}, \qquad \psi^{3+i}_{ab} = \delta^{ia} \delta^{4b} - \delta^{ib} \delta^{4a},
  \qquad \psi^{6+a}_{5b} = \delta^{ab}.
\end{equation}
Therefore, the 't Hooft symbols in \eq{thooft-matrix} are related to the psi-symbols by
\begin{equation}\label{two-symbol}
  \eta^i_{ab} = \psi^i_{ab} + \psi^{3+i}_{ab}, \qquad \bar{\eta}^i_{ab} = \psi^i_{ab} - \psi^{3+i}_{ab}.
\end{equation}
Then the above combination implies that
\begin{eqnarray} \label{su2-t+-}
  && T^i + T^{3+i} = \frac{1}{2} (\psi^i_{ab} + \psi^{3+i}_{ab})J_{ab} = \frac{1}{2} \eta^i_{ab} J_{ab} \in su(2)_+, \xx
  && T^i - T^{3+i} = \frac{1}{2} (\psi^i_{ab} - \psi^{3+i}_{ab})J_{ab} = \frac{1}{2} \bar{\eta}^i_{ab} J_{ab} \in su(2)_-,
\end{eqnarray}
where we used the definition \eq{linear-rel}. The coset generators are given by
\begin{equation}\label{coset-gen}
  T^{6+a} = \psi^{6+a}_{5b} J_{5b} = J_{5a} \in \mathfrak{k}.
\end{equation}
Thus the Cartan-Weyl basis for the root diagram in Fig. \ref{so5-root} can be classified as follows:
\begin{equation}\label{class}
  (H^1, A_\pm) \in su(2)_+, \qquad    (H^2, B_\pm) \in su(2)_-, \qquad (C_\pm, D_\pm) \in \mathfrak{k}.
\end{equation}
The decomposition of four-dimensional Einstein manifolds implies that $SU(2)_+$ Yang-Mills instantons live
in the vector space $su(2)_+$
and $SU(2)_-$ Yang-Mills anti-instantons live in the vector space $su(2)_-$.
The root diagram in Fig. \ref{so5-root} shows how each component in the five-dimensional Riemann curvature
tensors $R_{ABCD}$ deforms the instanton structure of four-dimensional Einstein manifolds.

Hence it is useful to decompose the five-dimensional Riemann curvature tensors in Eq. \eq{curv-eta}
according to the Lie algebra decomposition \eq{decom-so5}.
First the Riemann curvature tensors in \eq{5riem-abcd} are decomposed as
\begin{equation}\label{rdecom-abcd}
  R_{abcd} = \mathfrak{f}^{ij}_{(++)} \eta^i_{ab} \eta^j_{cd} + \mathfrak{f}^{ij}_{(+-)} \eta^i_{ab} \bar{\eta}^j_{cd}
  + \mathfrak{f}^{ij}_{(-+)}\bar{\eta}^i_{ab} \eta^j_{cd} + \mathfrak{f}^{ij}_{(--)} \bar{\eta}^i_{ab} \bar{\eta}^j_{cd},
\end{equation}
where
\begin{eqnarray} \label{def-f**}
  && \mathfrak{f}^{ij}_{(++)} \equiv \frac{1}{4} \big( \mathfrak{R}_{i,j} + \mathfrak{R}_{3+i, j}
  + \mathfrak{R}_{i,3+j} + \mathfrak{R}_{3+i, 3+j} \big), \xx
  && \mathfrak{f}^{ij}_{(+-)} \equiv \frac{1}{4} \big( \mathfrak{R}_{i,j} + \mathfrak{R}_{3+i, j}
  - \mathfrak{R}_{i,3+j} - \mathfrak{R}_{3+i, 3+j} \big), \xx
  && \mathfrak{f}^{ij}_{(-+)} \equiv \frac{1}{4} \big( \mathfrak{R}_{i,j} - \mathfrak{R}_{3+i, j}
  + \mathfrak{R}_{i,3+j} - \mathfrak{R}_{3+i, 3+j} \big), \xx
  && \mathfrak{f}^{ij}_{(--)} \equiv \frac{1}{4} \big( \mathfrak{R}_{i,j} - \mathfrak{R}_{3+i, j}
  - \mathfrak{R}_{i,3+j} + \mathfrak{R}_{3+i, 3+j} \big).
\end{eqnarray}
Explicitly, they are given by
\begin{eqnarray} \label{f**}
\mathfrak{f}^{ij}_{(++)}  &=& e^{\frac{1}{3} \phi} \left\{ f^{ij}_{(++)} - \kappa^2 e^\phi
\Big( \frac{3}{4} f^{(+)i} f^{(+)j} - \frac{1}{8} (f^{(+)k} f^{(+)k}- f^{(-)k} f^{(-)k}) \delta^{ij} \Big) \right. \xx
&& \qquad \left. + \frac{1}{24} \Big( \Delta \phi - \frac{1}{6} (\partial_a \phi)^2 \Big) \delta^{ij} \right\}, \xx
\mathfrak{f}^{ij}_{(+-)} &=& e^{\frac{1}{3} \phi} \left\{ f^{ij}_{(+-)} - \frac{3}{4}\kappa^2 e^\phi
f^{(+)i} f^{(-)j} + \frac{1}{24} \Big( D_a \partial_b \phi + \frac{1}{6} \partial_a \phi \partial_b \phi \Big)
\eta^i_{ac} \bar{\eta}^j_{bc}  \right\}, \\
\mathfrak{f}^{ij}_{(--)}  &=& e^{\frac{1}{3} \phi} \left\{ f^{ij}_{(--)} - \kappa^2 e^\phi
\Big( \frac{3}{4} f^{(-)i} f^{(-)j} + \frac{1}{8} (f^{(+)k} f^{(+)k}- f^{(-)k} f^{(-)k}) \delta^{ij} \Big) \right. \xx
&& \qquad \left. + \frac{1}{24} \Big( \Delta \phi - \frac{1}{6} (\partial_a \phi)^2 \Big) \delta^{ij} \right\}, \nonumber
\end{eqnarray}
where $ f^{ij}_{(**)}$ are the expansion coefficients of the four-dimensional Riemann curvature tensors \eq{decom-riem}
and we have introduced a similar decomposition for $U(1)$ field strengths
\begin{equation}\label{decom-fab}
  F_{ab} = f^{(+)i} \eta^i_{ab} + f^{(-)i}\bar{\eta}^i_{ab}.
\end{equation}
Note that
\begin{equation}\label{symm-fs}
\mathfrak{f}^{ij}_{(++)} = \mathfrak{f}^{ji}_{(++)}, \qquad   \mathfrak{f}^{ij}_{(--)} = \mathfrak{f}^{ji}_{(--)}, \qquad \mathfrak{f}^{ij}_{(+-)} = \mathfrak{f}^{ji}_{(-+)}
\end{equation}
due to the symmetry property \eq{symm-curv} and the Bianchi identity \eq{1st-bianchi} further requires
\begin{equation}\label{eq-trace}
  \mathfrak{f}^{ij}_{(++)} \delta^{ij} = \mathfrak{f}^{ij}_{(--)} \delta^{ij}.
\end{equation}
It is easy to check Eq. \eq{eq-trace} using the above results.\footnote{It can also be derived from Eq. \eq{iden-2}
by using the algebraic properties of the 't Hooft symbols in Appendix A:
\begin{eqnarray*}
\frac{1}{4} \varepsilon_{abcd5} R_{abcd} &=& \frac{1}{2} \left( \mathfrak{f}^{ij}_{(++)} \eta^i_{ab} \eta^j_{ab}
- \mathfrak{f}^{ij}_{(+-)} \eta^i_{ab} \bar{\eta}^j_{ab}
  + \mathfrak{f}^{ij}_{(-+)}\bar{\eta}^i_{ab} \eta^j_{ab} - \mathfrak{f}^{ij}_{(--)} \bar{\eta}^i_{ab} \bar{\eta}^j_{ab} \right) \\
  &=& 2 \left( \mathfrak{f}^{ij}_{(++)} \delta^{ij}
  - \mathfrak{f}^{ij}_{(--)} \delta^{ij} \right)=0.
\end{eqnarray*}
}
Using the decomposition \eq{decom-fab}, it is straightforward to calculate the $U(1)$ instanton density
\begin{eqnarray}\label{u1-idensity}
  \rho_{U(1)} &=& \frac{1}{64 \pi^2} \varepsilon^{abcd} F_{ab} F_{cd} \xx
  &=& \frac{1}{8 \pi^2} (f^{(+)k} f^{(+)k}- f^{(-)k} f^{(-)k}).
\end{eqnarray}

Using Eq. \eq{psi-comp}, the expansion \eq{curv-eta} for the Riemann tensors $R_{5abc}$ can be written as
\begin{eqnarray} \label{rdecom-5abc}
  R_{5abc} &=& \mathfrak{R}_{6+a, \mathbb{B}} \psi^\mathbb{B}_{bc} \xx
   &\equiv& \mathbb{F}^{(+)i}_{5a} \eta^i_{bc}  + \mathbb{F}^{(-)i}_{5a} \bar{\eta}^i_{bc},
\end{eqnarray}
where $\mathbb{F}^{(\pm)i}_{5a} = \frac{1}{2} \big(\mathfrak{R}_{6+a, i} \pm \mathfrak{R}_{6+a, 3+i} \big)$ are given by
\begin{eqnarray} \label{rf-5a}
  &&  \mathbb{F}^{(+)i}_{5a} = \frac{\kappa}{12} e^{\frac{5}{6} \phi} \Big( 6 D_a^{(+)} f^{(+)i}
  + 7 f^{(+)i} \partial_a \phi - 2 \varepsilon^{ijk} f^{(+)j} \eta^k_{ab} \partial_b \phi
  + f^{(-)j}\eta^i_{ac} \bar{\eta}^j_{bc} \partial_b \phi \Big), \xx
  &&  \mathbb{F}^{(-)i}_{5a} = \frac{\kappa}{12} e^{\frac{5}{6} \phi} \Big( 6 D_a^{(-)} f^{(-)i}
  + 7 f^{(-)i} \partial_a \phi - 2 \varepsilon^{ijk} f^{(-)j} \bar{\eta}^k_{ab} \partial_b \phi
  + f^{(+)j}\bar{\eta}^i_{ac} \eta^j_{bc} \partial_b \phi \Big),
\end{eqnarray}
and
\begin{equation}\label{cov-u1f}
  D_a^{(\pm)} f^{(\pm)i} = \partial_a f^{(\pm)i} - 2 \varepsilon^{ijk} A_a^{(\pm)j} f^{(\pm)k}.
\end{equation}
The expansion components $\mathbb{F}^{(\pm)i}_{5a}$ are not completely independent
due to the constraints \eq{5-bianchi}. It is straightforward to read off the constraints
using the Table \ref{table-d} in Appendix A that gives 4 relations from the first four rows,
so totally $20=24-4$ independent components remain.
The last line in the Table \ref{table-d} gives rise to the constraint \eq{eq-trace}.
Finally, we have
\begin{eqnarray} \label{rdecom-5a5b}
  R_{5a5b} &=& \mathfrak{R}_{6+a, 6+b} \equiv \mathfrak{R}_{ab} \xx
  &=& e^{\frac{1}{3} \phi} \left\{ \frac{\kappa^2}{4} e^\phi \Big( \big( f^{(+)i} f^{(+)i} + f^{(-)i} f^{(-)i} \big) \delta_{ab}
  + 2 f^{(+)i} f^{(-)j} \eta^i_{ac} \bar{\eta}^j_{bc} \Big) \right. \xx
  && \left. \qquad - \frac{2}{9} \partial_a \phi \partial_b \phi + \frac{1}{18} \partial_c \phi \partial_c \phi \delta_{ab}
  - \frac{1}{3} D_a \partial_b \phi \right\},
\end{eqnarray}
where $\mathfrak{R}_{ab} = \mathfrak{R}_{ba}$, so totally 10 components. Thus we recover the $50 = 20 + 20 + 10$ components
of Riemann curvature tensors in five dimensions.

\section{Five-dimensional Einstein Manifolds}

\begin{figure}
\centering
\includegraphics[width=0.3\textwidth]{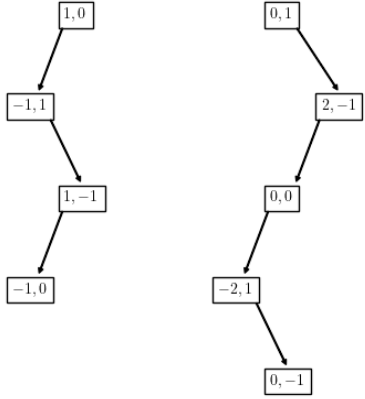}
\caption{The block weight diagrams of the fundamental representations of $sp(2) \cong so(5)$.}
\label{b-weight-diagram}
\end{figure}

It was shown in \eq{5lie-sp2} that $\psi^\mathbb{A}_{AB}$ defined in \eq{eta-symbol} provide the five-dimensional
representation of $sp(2) \cong so(5)$ Lie algebra. It is the irrep of $sp(2)$ corresponding to the highest weight
$\varpi_2 =(0,1)$ on the right-hand side in Fig. \ref{b-weight-diagram}.
It is well-known \cite{georgi-book} that a simple Lie algebra of rank $r$ possesses $r$ inequivalent fundamental irreps.
The two fundamental weights for the Lie algebra $sp(2) \cong so(5)$ are shown up in Fig. \ref{b-weight-diagram}.
The four-dimensional representation, corresponding to the highest weight
$\varpi_1 =(1,0)$ on the left-hand side in Fig. \ref{b-weight-diagram}, is the spinor representation of
$so(5)$ and the defining representation of $sp(2)$. In contrast, $\varpi_2 =(0,1)$ is the highest weight of
a five-dimensional representation of $sp(2) \cong so(5)$ Lie algebra.
It is easy to find the defining representation of $so(5)$ that is given by Eq. \eq{so5-defrep}.
There must exist a five-dimensional representation of $sp(2)$ defined by the fundamental weight
$\varpi_2 =(0,1)$. That is precisely provided by the psi-symbol \eq{eta-symbol}.

Therefore the expansion \eq{curv-eta} corresponds to the generalization of
the four-dimensional decomposition \eq{decom-riem} to the five-dimensional case.
The five-dimensional curvature tensors are not decomposed into some irreducible blocks because
the Lorentz group $SO(5) = Sp(2)/\mathbb{Z}_2$ is a simple group, unlike the four-dimensional case.
The Riemann curvature tensor belongs to the irrep of $SO(5) = Sp(2)/\mathbb{Z}_2$ given by
\begin{equation}\label{riem-50}
  R_{ABCD} \in \left(\mathbf{50} = {\tiny  \Yvcentermath1 \yng(2,2)} \right).
\end{equation}
Thus the five-dimensional Einstein manifold satisfying the equations of motion \eq{5ein-eq}
should take elements in the irrep \eq{riem-50}.
The expansion \eq{curv-eta} shows how these elements are organized according to the root structure in
Fig. \ref{sp2-root1} or Fig. \ref{so5-root}.
After the Kaluza-Klein compactification, the symmetry is reduced
to $SO(4) \times U(1)$. Then the Riemann curvature tensors in \eq{riem-50}
are decomposed according to the remaining symmetry \eq{symm-break} or more precisely Eq. \eq{decom-so5}.
This decomposition appears in Eqs. \eq{rdecom-abcd}, \eq{rdecom-5abc} and \eq{rdecom-5a5b}.
In particular, the components $f^{ij}_{(\pm\pm)}$ of four-dimensional Einstein manifolds appear
in the curvature tensor \eq{rdecom-abcd}. The instanton structure of Einstein manifolds is deformed
by the excitations of $U(1)$ gauge fields and a scalar field.
However, these deformations in the curvature tensor \eq{rdecom-abcd} are only done in the root directions
$A_\pm$ and $B_\pm$, corresponding to the $x$-axis and the $y$-axis, respectively, in the root diagram of
Fig. \ref{sp2-root1} or Fig. \ref{so5-root}. Therefore, if the mixed components $\mathfrak{f}^{ij}_{(+-)}
= \mathfrak{f}^{ji}_{(-+)}$ in the deformed curvature tensor $R_{abcd}$ identically vanish,
the instanton structure is still maintained despite the presence of $U(1)$ gauge fields and a scalar field, i.e.,
\begin{equation}\label{def-inst}
 \mathfrak{F}^{(\pm)i}_{ab} = \pm \frac{1}{2} \varepsilon_{abcd} \mathfrak{F}^{(\pm)i}_{cd},
\end{equation}
where $\mathfrak{F}^{(+)i}_{ab} \equiv \frac{1}{4} R_{abcd} \eta^i_{cd}$
and $\mathfrak{F}^{(-)i}_{ab} \equiv \frac{1}{4} R_{abcd} \bar{\eta}^i_{cd}$.
But generic excitations of four-dimensional fields break the instanton structure of
four-dimensional Einstein manifolds. Consequently, once the fifth dimension is opened so that
the Lorentz symmetry is enhanced to $SO(5)$,
all these deformations have to be organized into a single five-dimensional Einstein manifold.
So it may be interesting to look at some particular cases.

First, consider the case with $\phi = $ constant.
A caveat is that the condition $\phi = $ constant implies the unwanted result,
$\frac{1}{4} F_{\mu\nu} F^{\mu\nu} = \frac{\Lambda}{\kappa^2} e^{-\frac{4}{3} \phi}$,
from Eq. \eq{4eom-s}. In order to avoid this conclusion, one can proceed in the reverse order
by putting the condition $\phi = $ constant in the action \eq{kk-ehaction} or the ansatz \eq{kk-metric}
and varying the action afterwards \cite{kk-book}.
The constant scalar field can simply be removed by a field redefinition
and defining a four-dimensional cosmological constant
$\lambda \equiv \frac{3}{2} e^{- \frac{\kappa}{\sqrt{3}} \Phi} \Lambda$.
If we turn off the $U(1)$ gauge field, i.e. $A_\mu = 0$, we recover the four-dimensional Einstein manifolds
discussed in Sec. 2. We know well the instanton structure of four-manifolds in this case.
So let us turn on the $U(1)$ gauge field.
Note that the Einstein equation \eq{4eom-e} can be equivalently written as
\begin{equation}\label{ein-eq-5dim}
  R_{ab} - \frac{1}{2} \delta_{ab} R + \lambda e^{\frac{1}{3} \phi} \delta_{ab} = 0.
\end{equation}
Using the result \eq{ricci-decomp},
it is straightforward to reduce Eq. \eq{ein-eq-5dim} as the form \cite{oh-yang,joy-jhep}
\begin{equation}\label{eins-beq}
f^{ij}_{(++)} \delta^{ij} = f^{ij}_{(--)} \delta^{ij} = \frac{\lambda}{2},
\qquad f^{ij}_{(+-)} = 16 \pi G_4 f^{(+)i} f^{(-)j}.
\end{equation}
Since the Maxwell's equations are coming from the components of the Ricci tensor $R_{a5}=R_{5bab}$,
we can read off the expansion for the Maxwell's equations in the $sp(2) \cong so(5)$ basis from Eq. \eq{rdecom-5abc}:
\begin{eqnarray} \label{maxeq-sp2}
\eta^i_{ab} D_b^{(+)} f^{(+)i} + \bar{\eta}^i_{ab} D_b^{(-)} f^{(-)i} = 0
\end{eqnarray}
where the covariant derivatives are defined by Eq. \eq{cov-u1f}.

The structure in Eq. \eq{eins-beq} clearly shows that turning on $U(1)$ gauge fields
introduces a mixing of $SU(2)_+$ and $SU(2)_-$ sectors since the mixed part $f^{ij}_{(+-)}$ no longer vanishes.
Although the Riemann curvature tensor in this case does not satisfy the self-duality equation like Eq. \eq{def-inst},
the mixed part $f^{ij}_{(+-)}$ does not disturb the conformal and instanton structures of four-manifolds
since the Weyl tensor does not depend on the mixed part $f^{ij}_{(+-)} = f^{ji}_{(-+)}$ \cite{oh-yang}.
A bit mysterious aspect is that there is no effect in the four-dimensional Einstein equations \eq{eins-beq}
if only self-dual $(\mathrm{i.e.}, f^{(-)i} = 0)$ or anti-self-dual $(\mathrm{i.e.}, f^{(+)i} = 0)$ $U(1)$
gauge fields are turned on. This structure is due to the fact that the energy-momentum tensor
in Eq. \eq{em-decom} identically vanishes for self-dual or anti-self-dual gauge fields.
So one may conclude that the Einstein structure is infinitely degenerate in the sense
that one can add arbitrary self-dual or anti-self-dual $U(1)$ gauge fields without spoiling
the Einstein condition of a four-manifold. But the five-dimensional Einstein manifold secretly notices
the existence of such $U(1)$ instantons because the four-dimensional Maxwell's equations \eq{maxeq-sp2}
correspond to $R_{a5} = 0$ as shown by Eq. \eq{kk-52} and
they are nontrivial.\footnote{This kind of absurd insensitivity holds true even
when we consider a four-dimensional gravity coupled to $SU(2)$ Yang-Mills gauge theory \cite{oh-yang}.
The Einstein equations in this case are simply replaced by $f^{ij}_{(+-)}
= 16 \pi G_4 Tr \big( f^{(+)i} f^{(-)j} \big)$ where the trace is performed for the $SU(2)$ gauge group.
If $SU(2)$ gauge fields are Yang-Mills instantons whose equations are exactly the same as Eq. \eq{sdasd-eq},
$f^{ij}_{(+-)}$ again identically vanishes. Therefore, the Einstein structure is infinitely degenerate
even for the presence of Yang-Mills instantons. It is quite strange considering that the instanton equation of
Yang-Mills gauge fields is exactly the same as that of an Einstein manifold.
But a higher-dimensional Einstein manifold secretly notices
the existence of such Yang-Mills instantons as the five-dimensional case
because the four-dimensional Yang-Mills equations
are obtained by the Kaluza-Klein compactification of a higher-dimensional gravity \cite{cho-freund}.}

The deformed instanton structure defined by Eq. \eq{def-inst} does not allow a similar frigidity
as long as $U(1)$ gauge fields and a scalar field are active.
For simplicity, let us consider the case where only the scalar field is turned on but
$U(1)$ gauge fields are completely turned off.
Among the field configurations obeying the condition $\mathfrak{f}^{ij}_{(+-)} = 0$ which is equal
to the equations
\begin{equation}\label{mixed=0}
f^{ij}_{(+-)} = - \frac{1}{24} \Big( D_a \partial_b \phi + \frac{1}{6} \partial_a \phi \partial_b \phi \Big)
\eta^i_{ac} \bar{\eta}^j_{bc}
\end{equation}
with the covariant derivative $D_a \partial_b \phi = \partial_a \partial_b \phi - \omega_{cba} \partial_c \phi$,
the instanton Eq. \eq{def-inst} would not be affected by the scalar field if it satisfied
the equation
\begin{equation}\label{def-insens}
 \Delta \phi = \frac{1}{6} (\partial_a \phi)^2.
\end{equation}
But Eq. \eq{def-insens} cannot be satisfied for a nontrivial physical scalar field because
the left-hand side upon integration with a proper boundary (or asymptotic) condition
becomes negative while the right-hand side is positive-definite.
This implies that the instantons in four-dimensional Einstein manifolds are all connected by
activating the four-dimensional fields in the metric \eq{5-matrix}. Then it will be possible to bind
$SU(2)_+$ instantons and $SU(2)_-$ anti-instantons into a single multiplet of
the five-dimensional Lorentz group $SO(5) = Sp(2)/\mathbb{Z}_2$.
The unification of two independent instantons in a four-dimensional Einstein manifold would be clear
when looking at the root diagram in Fig. \ref{sp2-root1} or Fig. \ref{so5-root}.
In four dimensions, one can move only along the $x$-direction or the $y$-direction which
lies in the representation of $su(2)_+$ or $su(2)_-$, respectively, in Eq. \eq{class}.
These two classes cannot be mixed because the corresponding root vectors are orthogonal to each other.
But, in five dimensions, one can now move along the diagonal directions which correspond to the coset elements
in Eq. \eq{class}. Thus it will be possible to connect two kinds of instantons by exciting
four-dimensional fields coupled with $so(5) \cong sp(2)$ root vectors.
It will be left for future work to explicitly analyze the unification of four-dimensional Einstein manifolds
in five dimensions.

\section{Discussion}

There is a mysterious transition between Euclidean spaces and Minkowski (Lorentzian) spaces.
They are simply related by an analytic continuation $x^0 = - i x^4$, but it results in dramatic changes of physics.
In the Euclidean space, physical forces have the self-dual structure defined by Eq. \eq{sd-asd}.
The eigenspace of the self-dual structure is called instantons.
The 2-forms are important in Riemannian geometry because of their relation with the curvature tensor,
and this decomposition has a profound influence on the underlying geometry of four dimensions \cite{s-duality}.
And this separation is deeply related to the splitting of the Euclidean Lorentz group \eq{spin4-split}.
This correspondence is natural from the viewpoint of the Clifford isomorphism \cite{book-clifford} since
the 2-forms $\Omega^2$ in the exterior algebra is isomorphically related to the Lorentz generators
$J_{ab} = \frac{1}{4} [\gamma_a, \gamma_b]$ in the Clifford algebra.
After the Wick rotation $x^0 = - i x^4$, the physical forces no longer have the self-dual structure because
the Hodge $*$-operator satisfies $*^2 = - 1$.
Instead a novel structure emerges in the Minkowski (Lorentzian) space, the so-called causal structure.
A vector or more generally tensors have the causal structure depending on their signature;
timelike if $\|x\|^2 < 0$, spacelike if $\|x\|^2 > 0$ and lightlike if $\|x\|^2 =0$.
The causal structure in Euclidean spaces is trivial because always $\|x\|^2 > 0$
unless $x = 0$. Moreover, the Lorentz group $SO(3,1)$ becomes a simple group although it is a non-compact group.
The physical forces are no longer separated but they exert their influences according to the causality.
We wonder what the relationship between these two structures is.
A five-dimensional Lorentzian manifold may provide some clue for the question since time-independent solutions
can be classified by the four-dimensional self-dual structure and $SO(4) \subset SO(4,1)$.

Our formalism can also be applied to non-compact Riemannian manifolds with boundary.
However, in this case, it is necessary to include boundary terms to discuss the topological invariants
such as the Euler characteristic $\chi(M)$ and the Hirzebruch signature $\tau(M)$ \cite{egh-report,gib-pop,gib-haw,gib-3}.
These boundary terms introduce a mixing between $SU(2)_+$ gauge fields and $SU(2)_-$ gauge fields
in the topological invariants because the Lorentz symmetry $SO(4)\cong SU(2)_+ \otimes SU(2)_-/\mathbb{Z}_2$ is reduced
to $SO(3) \cong SU(2)/\mathbb{Z}_2$ on the boundary \cite{opy-jhep}.
The boundary symmetry $SU(2)$ corresponds to the diagonal element of $SU(2)_+ \otimes SU(2)_-$.
Thus the nice splitting between $SU(2)_+$ and $SU(2)_-$ factors is lost.
Furthermore, all known examples, at least, for gravitational instantons, imply \cite{gib-pop,gib-haw,gib-3}
that $\chi(M) = |\tau(M)| +1$. The reduction of the toplogical invariants is due to the reduction
of the Lorentz symmetry at the boundary.
The topologically inequivalent sector of instanton solutions is defined by the homotopy class of a map
from a three sphere at asymptotic infinity to the gauge group $G$
\begin{equation}\label{homotopy}
f: \mathbb{S}^3 \to G
\end{equation}
and the topological charge is given by an element of the homotopy group $\pi_3 (G)$ \cite{nakahara}.
Since the spin connection \eq{ga-split} can be viewed as gauge fields in $G=SO(4)$,
the topological sector of the $SO(4)$ gauge fields is given by the homotopy class
$\pi_3 (SO(4)) = \pi_3 (SU(2)_+ \times SU(2)_-) = \mathbb{Z}  \oplus \mathbb{Z}$.
Consequently, there are two independent topological charges, $\chi(M)$ and $\tau(M)$.
But, if a non-compact Riemannian manifold has a boundary, the Lorentz symmetry $SO(4)$
is reduced to $SO(3)$ due to the boundary and the homotopy class has to be defined by
the remaining symmetry, i.e. $\pi_3 (SO(3)) = \mathbb{Z}$.
This implies that the Euler characteristic $\chi(M)$ and the Hirzebruch signature $\tau(M)$ are
no longer independent but there must be some relation between them.
The relation  $\chi(M) = |\tau(M)| +1$ represents such a relationship.
It will be interesting to understand such a boundary effect from the Kaluza-Klein theory.
In particular, it is an interesting problem to include boundary terms in the action \eq{5-ehaction}
and understand a role of $U(1)$ gauge fields and a scalar field at the boundary.

The proton is a stable particle because it cannot decay to light leptons
due to the baryon number conservation.
However, in the GUT, a large simple group such as $SU(5)$ or $SO(10)$ contains quarks and leptons
in the same multiplet. Therefore it is possible for the proton to decay into a lepton
(a positron and two gamma ray photons) although its half-life is extremely long.
A similar instability of Einstein manifolds may appear in a five-dimensional gravity.
In five dimensions, the Lorentz group is $SO(5) \supset SO(4)$ which is a simple group.
Therefore $SU(2)_+$ instantons and $SU(2)_-$ anti-instantons must be embedded in the same multiplet
of $SO(5)$. The reason for the stability of a four-dimensional Einstein manifold is that
instantons and anti-instantons belong to different gauge groups as we intentionally indicated.
However, in five dimensions,
they belong to an irrep of the same simple group.
Then it is impossible to prevent these instantons from decaying each other.
The topological consideration also supports this conjecture.
In five dimensions, the Euler characteristic identically vanishes, $\chi = 0$.
It is a simple consequence of Poincar\'e duality that manifolds with an odd dimension
have vanishing Euler characteristic.
The Hirzebruch signature can also be defined only in multiples of 4 dimensions.
The homotopy consideration $f: \mathbb{S}^4 \to G$ similar to \eq{homotopy} also supports
this kind of triviality because $\pi_4 (SO(5)) = \mathbb{Z}_2$ (see Table 4.1 in \cite{nakahara}).
Thus there is no natural topological invariant to support the stability of a five-dimensional Einstein manifold.
If the fifth dimension is compactified with a sufficiently small radius,
the Lorentz symmetry \eq{symm-break} is reduced to $SO(4) \times U(1)$.
Then Einstein manifolds may recover their stability in four dimensions.
But this kind of instability in five dimensions may have appeared in early universe
and the universe would have stabilized in four dimensions through some similar mechanism in \cite{kky}.
It would be great if one could explain in this way why our universe have chosen the four-dimensional spacetime.

The standard Kaluza-Klein vacuum, $M \times \mathbb{S}^1$, is known to be unstable \cite{kk-bubble,kk-bubble1,kk-bubble2,bon-string}.
The instanton that mediates the decay is the five-dimensional Euclidean Schwarzschild solution
\begin{equation}\label{kk-bubble}
  ds^2 = \frac{dr^2}{1-\frac{R^2}{r^2}} + r^2 d\Omega_3^2 + \left(1-\frac{R^2}{r^2}\right) d \chi^2
\end{equation}
where $d\Omega_3^2$ is the metric on the unit three-sphere and $\chi$ is the coordinate on the Kaluza-Klein
circle. The five-dimensional solution \eq{kk-bubble} is a bounce which describes the decay of
the Kaluza-Klein vacuum and is a topology changing process.
In four dimensions, the Euclidean Schwarzschild solution \cite{euc-bh} is Ricci-flat and
it consists of an $SU(2)_+$ instanton and an $SU(2)_-$ anti-instanton.
The solution is semi-classically stable since it carries nontrivial topological invariants,
$\chi(M) = 1+1 = 2$ and $\tau(M) = 0$ \cite{egh-report,opy-jhep}.
However, if the four-dimensional Schwarzschild solution
is lifted to the five-dimensional solution \eq{kk-bubble},
it was shown in \cite{kk-bubble} that a nonperturbative instability of the ground state is developed.
Therefore, it will be interesting to investigate the nature of
instability in \cite{kk-bubble,kk-bubble1,kk-bubble2,bon-string} from the perspective we have discussed above.
Any progress in this direction will be reported.

\section*{Acknowledgments}

We thank the anonymous referee for a careful reading of the manuscript and helpful criticisms with some explanations.
HSY was supported in part by the National Research Foundation of Korea (NRF)
with grant numbers NRF-2018R1D1A1B0705011314.
We also appreciate APCTP for its hospitality during completion of this work.


\appendix

\section{$SO(4) \cong SU(2)_+ \otimes SU(2)_-/\mathbb{Z}_2$ and $SO(5) = Sp(2)/\mathbb{Z}_2$}

The defining representation of the Lie algebra $so(n)$ is
$$ so(n) = \{ M| M \in \mathfrak{gl}(n, \mathbb{R}) \; \mathrm{such \, that} \; M^T = -M \}.$$
We take six generators of the Lie algebra $so(4)$ as
\begin{equation}\label{ab-so4}
  (X_i)_{ab} = - \varepsilon_{iab4}, \qquad (Y_i)_{ab} = - (\delta_{ai} \delta_{b4} - \delta_{bi} \delta_{a4} ),
\end{equation}
where $i=1,2,3; \; a,b = 1, \cdots, 4$ and the Levi-Civita tensor is normalized as $\varepsilon_{1234} =1$.
Two sets $(X_i, Y_i)$ satisfy the commutation relations
$$[X_i, X_j] = \varepsilon_{ijk} X_k, \quad [Y_i, Y_j] = \varepsilon_{ijk} X_k, \quad [X_i, Y_j] = \varepsilon_{ijk} Y_k$$
where $\varepsilon_{ijk}= \varepsilon_{ijk4}$.
It is convenient to define a new set of generators as
\begin{equation}\label{t+-}
  \tau^\pm_i = - \frac{1}{2} (X_i \pm Y_i).
\end{equation}
Then $\tau_i^\pm$ satisfy $so(3)$ or $su(2)$ Lie algebra, separately,
\begin{equation}\label{t-lie}
  [\tau^\pm_i, \tau^\pm_j ]  = - \varepsilon_{ijk} \tau^\pm_k, \qquad [\tau^\pm_i, \tau^\mp_j ] = 0.
\end{equation}
Hence the Lie algebra $so(4)$ is a direct sum of two independent $so(3)$ or $su(2)$ Lie algebras:
\begin{equation}\label{4-lie-iso}
so(4) \cong so(3)_+ \oplus so(3)_- \cong su(2)_+ \oplus su(2)_-.
\end{equation}
Since the direct sum of Lie algebras corresponds to the direct product of Lie groups
and the universal covering group of $SO(3)$ is $SU(2)$, we get the group isomorphism \cite{book-wybo}
\begin{equation}\label{4-group-iso}
  SO(4) \cong SU(2)_+ \otimes SU(2)_-/\mathbb{Z}_2.
\end{equation}

One can identify the components of two families of $4 \times 4$ matrices $\tau_i^\pm$ from Eq. \eq{ab-so4}:
\begin{eqnarray} \label{thooft-matrix}
&& 2 [\tau^i_+]_{ab} \equiv \eta^i_{ab}= {\varepsilon}^{iab4} + (\delta^{ia}\delta^{4b}
- \delta^{ib}\delta^{4a}), \xx
&& 2 [\tau^i_-]_{ab} \equiv {\overline \eta}^i_{ab} = {\varepsilon}^{iab4} - (\delta^{ia}\delta^{4b}
- \delta^{ib}\delta^{4a}).
\end{eqnarray}
Explicitly, they are given by \cite{joy-jhep,opy-jhep}
\begin{eqnarray} \label{t+}
&& \tau^{1}_+ = \frac{1}{2} \begin{pmatrix}
      0 & 0 & 0 & 1 \\
      0 & 0 & 1 & 0 \\
      0 & -1 & 0 & 0 \\
      -1 & 0 & 0 & 0 \\
             \end{pmatrix}, \;\;
  \tau^{2}_+ = \frac{1}{2} \begin{pmatrix}
      0 & 0 & -1 & 0 \\
      0 & 0 & 0 & 1 \\
      1 & 0 & 0 & 0 \\
      0 & -1 & 0 & 0 \\
    \end{pmatrix}, \;\;
  \tau^{3}_+ = \frac{1}{2} \begin{pmatrix}
      0 & 1 & 0 & 0 \\
      -1 & 0 & 0 & 0 \\
      0 & 0 & 0 & 1 \\
      0 & 0 & -1 & 0 \\
    \end{pmatrix}, \qquad \\
\label{t-}
&& \tau^{1}_- = \frac{1}{2} \begin{pmatrix}
      0 & 0 & 0 & -1 \\
      0 & 0 & 1 & 0 \\
      0 & -1 & 0 & 0 \\
      1 & 0 & 0 & 0 \\
    \end{pmatrix}, \;\;
  \tau^{2}_- = \frac{1}{2} \begin{pmatrix}
      0 & 0 & -1 & 0 \\
      0 & 0 & 0 & -1 \\
      1 & 0 & 0 & 0 \\
      0 & 1 & 0 & 0 \\
    \end{pmatrix}, \;\;
  \tau^{3}_- = \frac{1}{2} \begin{pmatrix}
      0 & 1 & 0 & 0 \\
      -1 & 0 & 0 & 0 \\
      0 & 0 & 0 & -1 \\
      0 & 0 & 1 & 0 \\
    \end{pmatrix}.
\end{eqnarray}
The matrices in (\ref{thooft-matrix}) provide two independent spin $s=\frac{3}{2}$ representations
of $su(2)$ Lie algebra.
The so-called 't Hooft symbols defined by Eq. \eq{thooft-matrix} satisfy the following relations \cite{joy-jhep,opy-jhep}
\begin{eqnarray} \label{self-eta}
&& \eta^{(\pm)i}_{ab} = \pm \frac{1}{2} {\varepsilon_{ab}}^{cd}
\eta^{(\pm)i}_{cd}, \\
\label{proj-eta}
&& \eta^{(\pm)i}_{ab} \eta^{(\pm)i}_{cd} =
\delta_{ac}\delta_{bd}
-\delta_{ad}\delta_{bc} \pm \varepsilon_{abcd}, \\
\label{self-eigen}
&& \varepsilon_{abcd} \eta^{(\pm)i}_{de} = \mp (
\delta_{ec} \eta^{(\pm)i}_{ab} + \delta_{ea} \eta^{(\pm)i}_{bc} -
\delta_{eb} \eta^{(\pm)i}_{ac} ), \\
\label{eta-etabar}
&& \eta^{(\pm)i}_{ab} \eta^{(\mp)j}_{ab}=0, \\
\label{eta^2}
&& \eta^{(\pm)i}_{ac}\eta^{(\pm)j}_{bc} =\delta^{ij}\delta_{ab} +
\varepsilon^{ijk}\eta^{(\pm)k}_{ab}, \\
\label{eta-ex}
&& \eta^{(\pm)i}_{ac}\eta^{(\mp)j}_{bc} =
\eta^{(\pm)i}_{bc}\eta^{(\mp)j}_{ac}, \\
\label{eta-o4-algebra}
&& \varepsilon^{ijk} \eta^{(\pm)j}_{ab} \eta^{(\pm)k}_{cd} =
    \delta_{ac} \eta^{(\pm)i}_{bd} - \delta_{ad} \eta^{(\pm)i}_{bc}
    - \delta_{bc} \eta^{(\pm)i}_{ad} + \delta_{bd} \eta^{(\pm)i}_{ac},
\end{eqnarray}
where $\eta^{(+)i}_{ab} \equiv \eta^i_{ab}$ and $\eta^{(-)i}_{ab} \equiv {\overline
\eta}^i_{ab}$.

Now we consider the $so(5)$ Lie algebra.
The five-dimensional gamma matrices $\gamma_A = (\gamma_a, \gamma_5), \; A=1, \cdots, 5$, are given by
\begin{equation}\label{5-gamma}
  \gamma_a = \left(
               \begin{array}{cc}
                 0 & \sigma^a \\
                 \bar{\sigma}^a & 0 \\
               \end{array}
             \right), \qquad \gamma_5 = -\gamma_1 \gamma_2 \gamma_3 \gamma_4 = \left(
                                                                        \begin{array}{cc}
                                                                          \mathbf{1}_2 & 0 \\
                                                                          0 & - \mathbf{1}_2 \\
                                                                        \end{array}
                                                                      \right),
\end{equation}
where $\sigma^a = (i \tau^i, \mathbf{1}_2 )$ and $\bar{\sigma}^a = (-i \tau^i, \mathbf{1}_2 ) = (\sigma^a)^\dagger$
with $\tau^i$ the Pauli matrices.
They satisfy the Dirac algebra
\begin{equation}\label{5-dirac}
  \{\gamma_A, \gamma_B \} = 2 \delta_{AB}.
\end{equation}
Then the Lorentz generators of $so(5)$ Lie algebra are defined by
\begin{equation}\label{5-gen}
  J_{AB} = \frac{1}{4} [\gamma_A, \gamma_B].
\end{equation}
One can see that the four-dimensional Lorentz algebra generated by $J_{ab}$ obeys the chiral representation
(see Eq. \eq{tso5-dec}) whose generators are given by
\begin{equation}\label{4-chiral}
J_{ab}^{(\pm)} = \frac{1}{2} (1 \pm \gamma_5) J_{ab}
\end{equation}
and
\begin{equation}\label{2-chiral}
J_{ab}^{(+)} = \frac{i}{2} \eta^i_{ab} \tau^i \in su(2)_+, \qquad
J_{ab}^{(-)} = \frac{i}{2} \bar{\eta}^i_{ab} \tau^i \in su(2)_-.
\end{equation}
The 't Hooft symbols in Eq. \eq{thooft-matrix} are obtained by
\begin{equation}\label{chiral-thooft}
  \eta^i_{ab} = -i Tr (  J_{ab}^{(+)} \tau^i ), \qquad \bar{\eta}^i_{ab} = -i Tr (  J_{ab}^{(-)} \tau^i ).
\end{equation}
The chiral generators in Eq. \eq{4-chiral} independently satisfy the four-dimensional Lorentz algebra that verifies
the Lie algebra isomorphism \eq{4-lie-iso}.

It is easy to check the Lie algebra isomorphism $sp(2) \cong so(5)$ using the identification \eq{tab-45}.
The matrices $T^\mathbb{A}$ in \eq{tabcd} are anti-Hermitian, i.e.,
$(T^\mathbb{A})^\dagger = - T^\mathbb{A}$ and obey the relation
\begin{equation}\label{sp2-cond}
 (T^\mathbb{A})^T J + J T^\mathbb{A} = 0
\end{equation}
where $J = i (\tau^2 \otimes \mathbf{1}_2) = i \left(
                                   \begin{array}{cc}
                                     \tau^2 & 0 \\
                                     0 & \tau^2 \\
                                   \end{array}
                                 \right)$ is the symplectic matrix.
The relation \eq{sp2-cond} implies that the $4 \times 4$ matrices $T^\mathbb{A}$ in \eq{tabcd} are the Lie algebra generators
of $sp(2)$. Indeed they satisfy the $sp(2)$ Lie algebra
\begin{equation}\label{sp2-lie}
  [T^\mathbb{A}, T^\mathbb{B}] = -f^{\mathbb{A} \mathbb{B} \mathbb{C}} T^\mathbb{C}
\end{equation}
where $f^{\mathbb{A}\mathbb{B}\mathbb{C}}$ are totally antisymmetric structure constants.
Their nonzero components are listed below
\begin{equation}\label{sp2-str-const}
  f^{ijk} = f^{(3+i)(3+j) k} = f^{(6+i)(6+j) k} = \varepsilon^{ijk}, \qquad f^{(3+i)(6+j) 10} = \delta^{ij}
\end{equation}
that can be read off from Eq. \eq{sp2-comm}.
Thus we establish the Lie algebra isomorphism $sp(2) \cong so(5)$.

The $so(5)$ generators in \eq{5-gen} also obey the anti-commutation relation
\begin{equation}\label{anti-so5}
  \{J_{AB}, J_{CD} \} = - \frac{1}{2} (\delta_{AC} \delta_{BD} - \delta_{AD} \delta_{BC}) \mathbf{1}_4
  - \frac{1}{2} \varepsilon_{ABCDE} \gamma_E.
\end{equation}
The corresponding anti-commutation relations for the $sp(2)$ generators in \eq{tabcd} read as
\begin{equation}\label{anti-sp5}
  \{T^\mathbb{A}, T^\mathbb{B} \} = - \frac{1}{2} \delta^{\mathbb{A}\mathbb{B}} \mathbf{1}_4
  - \frac{1}{2} d^{\mathbb{A}\mathbb{B} C} \widetilde{T}^C,
\end{equation}
where $\widetilde{T}^C$ do not belong to the set of $sp(2)$ generators.
Indeed they are given by $\widetilde{T}^A = \gamma_A$ and correspond to the second term on the right-hand side
of Eq. \eq{anti-so5}. The non-vanishing components of $d^{\mathbb{A}\mathbb{B} C}$ are listed in Table \ref{table-d}.
\begin{table}[ht]
\centering
\begin{tabular}{|c|c|c|}
  \hline
  $-(1,10,1)$ & $(5,9,1)$ & $-(6,8,1)$ \\
  \hline
  $- (2,10,2)$ & $-(4,9,2)$ & $(6,7,2)$ \\
  \hline
  $-(3,10,3)$ & $(4,8,3)$ & $-(5,7,3)$ \\
  \hline
  $(1,7,4)$ & $(2,8,4)$ & $(3,9,4)$ \\
  \hline
  $(1,4,5)$ & $(2,5,5)$ & $(3,6,5)$ \\
  \hline
\end{tabular}
\caption{$\pm(\mathbb{A},\mathbb{B}, C) \equiv \pm d^{\mathbb{A}\mathbb{B} C} = 1.$}
\label{table-d}
\end{table}
Then one can deduce the product of $so(5)$ and $sp(2)$ generators
\begin{eqnarray}\label{prod-gen1}
&& J_{AB} J_{CD} = - \frac{1}{4} (\delta_{AC} \delta_{BD} - \delta_{AD} \delta_{BC}) \mathbf{1}_4
   - \frac{1}{2} ( \delta_{AC} J_{BD} - \delta_{AD} J_{BC} - \delta_{BC} J_{AD} + \delta_{BD} J_{AC}) \xx
   && \hspace{2cm}   - \frac{1}{4} \varepsilon_{ABCDE} \gamma_E, \\
\label{prod-gen2}
&& T^\mathbb{A} T^\mathbb{B} = - \frac{1}{4} \delta^{\mathbb{A}\mathbb{B}} \mathbf{1}_4
  - \frac{1}{2} f^{\mathbb{A} \mathbb{B} \mathbb{C}} T^\mathbb{C}
  - \frac{1}{4} d^{\mathbb{A}\mathbb{B} C} \widetilde{T}^C.
\end{eqnarray}
Using the linear relation \eq{linear-rel}, let us write the products in Eqs. \eq{prod-gen1} and \eq{prod-gen2}
as the form
\begin{eqnarray}
   &(I)&  J_{AB} J_{CD} =  \psi^\mathbb{A}_{AB} \psi^\mathbb{B}_{CD} T^\mathbb{A} T^\mathbb{B}, \\
   &(II)&  T^\mathbb{A} T^\mathbb{B} = \frac{1}{4} \psi^\mathbb{A}_{AB} \psi^\mathbb{B}_{CD} J_{AB} J_{CD}.
\end{eqnarray}
Applying the relations in Eqs. \eq{prod-gen1} and \eq{prod-gen2} on both sides in the products $(I)$ and $(II)$
leads to useful algebraic relations for the psi-symbols \eq{eta-symbol}.
From the product $(I)$, we get
\begin{eqnarray} \label{5eta-1}
  &&  \psi^\mathbb{A}_{AB} \psi^\mathbb{A}_{CD} = \delta_{AC} \delta_{BD} - \delta_{AD} \delta_{BC}, \\
  \label{5eta-2}
  &&  f^{\mathbb{A} \mathbb{B} \mathbb{C}} \psi^\mathbb{B}_{AB} \psi^\mathbb{C}_{CD}
  = \delta_{AC} \psi^\mathbb{A}_{BD} - \delta_{AD} \psi^\mathbb{A}_{BC} - \delta_{BC} \psi^\mathbb{A}_{AD}
  + \delta_{BD} \psi^\mathbb{A}_{AC},
\end{eqnarray}
where $Tr (\gamma_E T^\mathbb{A}) = Tr ( \widetilde{T}^C T^\mathbb{A}) = 0$ were used.
Similarly, from the product $(II)$, we get
\begin{eqnarray} \label{5eta-3}
  &&  \psi^\mathbb{A}_{AB} \psi^\mathbb{B}_{AB} = 2 \delta^{\mathbb{A}\mathbb{B}}, \\
  \label{5eta-4}
  &&  f^{\mathbb{A} \mathbb{B} \mathbb{C}} \psi^\mathbb{C}_{AB}
  = \psi^\mathbb{A}_{AC} \psi^\mathbb{B}_{BC} - \psi^\mathbb{A}_{BC} \psi^\mathbb{B}_{AC},
\end{eqnarray}
where $Tr (\gamma_E J_{AB}) = Tr ( \widetilde{T}^C J_{AB}) = 0$ were used.
The above relations are analogous to those in Eqs. \eq{self-eta}-\eq{eta-o4-algebra}.
Actually those identities have been derived by applying a similar technique to Eq. \eq{2-chiral}.
If we define $5 \times 5$ matrices by
\begin{equation}\label{5-sp2}
  (T^\mathbb{A})_{AB} \equiv \psi^\mathbb{A}_{AB},
\end{equation}
Eq. \eq{5eta-4} reduces to the commutation relations
\begin{equation}\label{5lie-sp2}
  [T^\mathbb{A}, T^\mathbb{B}] = - f^{\mathbb{A} \mathbb{B} \mathbb{C}} T^\mathbb{C},
\end{equation}
while Eq. \eq{5eta-3} gives us the trace $Tr( T^\mathbb{A} T^\mathbb{B} ) = - 2 \delta^{\mathbb{A}\mathbb{B}}$.
Therefore the generators in \eq{5-sp2} provide the five-dimensional representation of $sp(2)$ Lie algebra
which is isomorphic to the defining representation of $so(5)$ Lie algebra with generators given by
\begin{equation}\label{so5-defrep}
  (J_{AB})_{CD} = \delta_{AC} \delta_{BD} - \delta_{AD} \delta_{BC}.
\end{equation}
The relation \eq{5eta-1} corresponds to the Fierz identity for the $sp(2)$ Lie algebra generators in \eq{5-sp2}
and the identity \eq{5eta-2} can be transformed into Eq. \eq{5eta-4} by using the trace \eq{5eta-3}
or vice versa by using the Fierz identity \eq{5eta-1}.

\section{Kaluza-Klein gravity}

On a five-dimensional Riemannian manifold $N$, the spin connection $\Omega = \frac{1}{2} \Omega_{AB} J^{AB}
= \frac{1}{2} \Omega_{AB M} J^{AB} dX^M$ constitutes an $SO(5)$ gauge field with respect to the local
$SO(5)$ rotations
$$ \Omega \to \Omega'= \Lambda \Omega \Lambda^{-1} + \Lambda d \Lambda^{-1} $$
where $\Lambda = \exp (\frac{1}{2} \lambda_{AB}(X) J^{AB} ) \in SO(5)$ and $J^{AB}$ are $so(5)$ Lorentz generators
in \eq{5-gen}.
Then the covariant derivatives for the vectors $E_A$ and $E^A$ are defined by
\begin{eqnarray*}
  && D_M \mathcal{E}_A = \partial_M \mathcal{E}_A - {\Omega^B}_{A M} \mathcal{E}_B,   \\
  && D_M E^A = \partial_M E^A + {\Omega^A}_{B M} E^B.
\end{eqnarray*}

The connection one-forms ${\Omega^A}_B = {\Omega^A}_{B M} dX^M$ satisfy
the Cartan's structure equations \cite{egh-report,nakahara}
\begin{eqnarray} \label{steq-torsion}
  T^A &=& dE^A + {\Omega^A}_B \wedge E^B, \\
  \label{steq-curvature}
  {R^A}_B  &=& d {\Omega^A}_B + {\Omega^A}_C \wedge {\Omega^C}_B,
\end{eqnarray}
where $T^A$ are the torsion two-forms and ${R^A}_B$ are the curvature two-forms.
We impose the torsion-free condition, $T^A_{MN} = D_M E_N^A - D_N E_M^A =0$, to recover the standard content
of general relativity which determines $\Omega_M$ as
\begin{equation}\label{spin-torsion}
  \Omega_{ABC} = \Omega_{ABM} \mathcal{E}_C^M = \frac{1}{2} (f_{BCA} + f_{CAB} - f_{ABC})
\end{equation}
where $f_{ABC}$ are the structure functions defined by
\begin{equation}\label{str-vector}
  [\mathcal{E}_A, \mathcal{E}_B] = - {f_{AB}}^{C} \mathcal{E}_C
\end{equation}
or its dual equations
\begin{equation}\label{str-covector}
  dE^A = \frac{1}{2} {f_{BC}}^{A} E^B \wedge E^C.
\end{equation}

In order to formulate gravity as a gauge theory of local Lorentz symmetry,
it is necessary to introduce a local basis
for the Kaluza-Klein geometry \eq{kk-metric}:
\begin{equation}\label{5-viel}
  E^A = (E^a, E^5) = ( e^{-\frac{1}{6} \phi} e^a,  e^{\frac{1}{3} \phi} e^5 )
\end{equation}
where
\begin{equation}\label{4-vier}
 ds_4^2 = e^a \otimes e^a = g_{\mu\nu} (x) dx^\mu dx^\nu
\end{equation}
and
\begin{equation}\label{5th-e}
  e^5 = dx^5 + \kappa A_\mu (x) dx^\mu.
\end{equation}
By solving the torsion-free condition $T^A=0$ in Eq. \eq{steq-torsion},
one can determine the spin connections as
\begin{eqnarray} \label{kk-spin}
  \Omega_{ab} &=& \omega_{ab} - \frac{1}{6}e^{\frac{1}{6} \phi} (\partial_b \phi E^a - \partial_a \phi E^b)
  - \frac{\kappa}{2} e^{\frac{2}{3} \phi} F_{ab} E^5, \\
  \Omega_{a5} &=& - \frac{1}{3} e^{\frac{1}{6} \phi} \partial_a \phi E^5
  - \frac{\kappa}{2} e^{\frac{2}{3} \phi} F_{ab} E^b,
\end{eqnarray}
where $\omega_{ab}$ is the four-dimensional spin connection for the local frames in \eq{4-vier}
and $\partial_a \equiv \mathfrak{e}_a = \mathfrak{e}_a^\mu (x) \partial_\mu \in \Gamma(TM)$ are orthonormal tangent vectors
dual to the covectors $e^a$, i.e. $\langle e^a, \mathfrak{e}_b \rangle = \delta^a_b$.
In particular, the exterior derivative acting on $M \times \mathbb{S}^1$ is defined by
$d = dx^\mu \partial_\mu = e^a \partial_a$ since we have assumed the cylinder condition (i.e., no dependence on $x^5$).

After a little algebra, the curvature two-forms for the Kaluza-Klein geometry \eq{kk-metric} can be determined
by the structure equation \eq{steq-curvature}:
\begin{eqnarray}\label{5curv-ab}
  R_{ab} &=&  \leftidx{^{(4)}}{R}_{ab} - \frac{\kappa^2}{4} e^\phi (F_{ab} F_{cd} + F_{ac} F_{bd}) e^c \wedge e^d
  - \frac{\kappa}{2} e^\phi D_c F_{ab} e^c \wedge e^5 \xx
  && - \frac{\kappa}{4} e^\phi ( 2F_{ab} \partial_c \phi - F_{bc} \partial_a \phi + F_{ac} \partial_b \phi) e^c \wedge e^5 \xx
  && + \frac{\kappa}{12} e^\phi ( F_{ac} \partial_c \phi  e^b \wedge e^5 - F_{bc} \partial_c \phi  e^a \wedge e^5) \xx
  && + \frac{1}{36} (\partial_b \phi \partial_c \phi  e^a \wedge e^c - \partial_a \phi  \partial_c \phi e^b \wedge e^c
  - \partial_c \phi  \partial_c \phi e^a \wedge e^b ) \xx
  && + \frac{1}{6} (D_c \partial_a \phi  e^c \wedge e^b - D_c \partial_b \phi e^c \wedge e^a), \\
  \label{5curv-5a}
  R_{5a} &=& \frac{\kappa^2}{4} e^{\frac{3}{2}\phi} F_{ab} F_{bc}  e^c \wedge e^5
  + \frac{\kappa}{2} e^{\frac{1}{2}\phi} D_c F_{ab} e^c \wedge e^b \xx
  && + \frac{\kappa}{4} e^{\frac{1}{2}\phi} ( F_{bc} \partial_a \phi - F_{ab} \partial_c \phi ) e^b \wedge e^c
  - \frac{\kappa}{12} e^{\frac{1}{2}\phi} F_{bc} \partial_b \phi e^a \wedge e^c \xx
  && + \frac{1}{9} e^{\frac{1}{2}\phi} (2 \partial_a \phi \partial_b \phi + 3 D_b \partial_a \phi ) e^b \wedge e^5
  - \frac{1}{18} e^{\frac{1}{2}\phi} \partial_b \phi \partial_b \phi  e^a \wedge e^5.
\end{eqnarray}
Here $\leftidx{^{(4)}}{R}_{ab}$ is the curvature two-form determined by the four-dimensional metric \eq{4-vier}
and the covariant derivatives are defined by
\begin{eqnarray*}
  &&  D_a \partial_b \phi = \partial_a \partial_b \phi - \omega_{cba} \partial_c \phi, \\
  && D_c F_{ab} = \partial_c F_{ab} - \omega_{dac} F_{db} - \omega_{dbc}  F_{ad}.
\end{eqnarray*}
Note that the derivations $\partial_a \equiv \mathfrak{e}_a$ do not commute
but they satisfy the structure equation similar to Eq. \eq{str-vector},
\begin{equation}\label{4d-steq}
  [\partial_a, \partial_b] = - {\mathfrak{f}_{ab}}^c \partial_c.
\end{equation}

The curvature two-forms above have the following expansion in the basis $(e^c \wedge e^d, e^c \wedge e^5)$
\begin{eqnarray*}
  R_{ab} &=& \frac{1}{2} e^{-\frac{1}{3}\phi} R_{abcd} e^c \wedge e^d
  + e^{\frac{1}{6}\phi} R_{abc5} e^c \wedge e^5, \\
  R_{5a} &=& \frac{1}{2} e^{-\frac{1}{3}\phi} R_{5abc} e^b \wedge e^c
  + e^{\frac{1}{6}\phi} R_{5a5b} e^5 \wedge e^b.
\end{eqnarray*}
Therefore one can read off the components of Riemann curvature tensors from Eqs. \eq{5curv-ab} and \eq{5curv-5a}:
\begin{eqnarray}\label{5riem-abcd}
   R_{abcd} &=& e^{\frac{1}{3}\phi} \left\{ \leftidx{^{(4)}}{R}_{abcd}
  - \frac{\kappa^2}{4} e^\phi (2 F_{ab} F_{cd} + F_{ac} F_{bd} - F_{ad} F_{bc} ) \right. \xx
  && + \frac{1}{36} \Big( \partial_a \phi \partial_c \phi \delta_{bd} - \partial_a \phi  \partial_d \phi \delta_{bc}
 - \partial_b \phi \partial_c \phi \delta_{ad} + \partial_b \phi  \partial_d \phi \delta_{ac}
  - \partial_e \phi  \partial_e \phi (\delta_{ac} \delta_{bd} - \delta_{ad}\delta_{bc}) \Big) \xx
  && \left. + \frac{1}{6} (D_c \partial_a \phi \delta_{bd} - D_d \partial_a \phi \delta_{bc}
  - D_c \partial_b \phi \delta_{ad} + D_d \partial_b \phi \delta_{ac} ) \right\}, \\
\label{5riem-abc5}
 R_{abc5} &=&  - \frac{1}{2} \kappa e^{\frac{5}{6}\phi} \left\{F_{ab} \partial_c \phi
  - \frac{1}{2} F_{bc} \partial_a \phi + \frac{1}{2} F_{ac} \partial_b \phi
 - \frac{1}{6} (F_{ad}  \delta_{bc} - F_{bd} \delta_{ac} ) \partial_d \phi + D_c F_{ab} \right\}, \\
\label{5riem-5a5b}
  R_{5a5b} &=& e^{\frac{1}{3}\phi} \left\{ \frac{\kappa^2}{4} e^\phi F_{ac} F_{bc}
  - \frac{2}{9} \partial_a \phi \partial_b \phi + \frac{1}{18} \partial_c \phi \partial_c \phi \delta_{ab}
  - \frac{1}{3} D_b \partial_a \phi \right\}.
\end{eqnarray}
Note that
\begin{equation}\label{comm-cov}
  D_a \partial_b \phi - D_b \partial_a \phi = ( - \mathfrak{f}_{abc} + \omega_{cab} - \omega_{cba} ) \partial_c \phi = 0,
\end{equation}
because $\mathfrak{f}_{abc} = \omega_{cab} - \omega_{cba}$. Therefore, $R_{5a5b} = R_{5b5a}$ as it should be.

Now it is easy to determine the Ricci tensors and the Ricci scalar using the above results:
\begin{eqnarray}\label{kk-51}
R_{ab} &=& R_{cacb} + R_{5a5b} \xx
         &=&  e^{\frac{1}{3}\phi} \Big( \leftidx{^{(4)}}{R}_{ab}
         - \frac{\kappa^2}{2} e^\phi F_{ac} F_{bc} - \frac{1}{6} \partial_a \phi \partial_b \phi
         + \frac{1}{6} D_c \partial_c \phi \delta_{ab} \Big), \\
\label{kk-52}
R_{a5} &=& R_{bab5} \xx
        &=& \frac{\kappa}{2} e^{\frac{5}{6}\phi} (F_{ab} \partial_b \phi + D_b F_{ab} ), \\
\label{kk-53}
R_{55} &=& R_{a5a5} \xx
        &=& e^{\frac{1}{3}\phi} \Big( \frac{\kappa^2}{4} e^\phi F_{ab} F_{ab}
        - \frac{1}{3} D_a \partial_a \phi \Big),   \\
\label{kk-54}
R &=& R_{ab} \delta^{ab} + R_{55} \xx
&=& e^{\frac{1}{3}\phi} \Big( \leftidx{^{(4)}}{R}
         - \frac{\kappa^2}{4} e^\phi F_{ab} F_{ab} - \frac{1}{6} \partial_a \phi \partial_a \phi
         + \frac{1}{3} D_a \partial_a \phi \Big),
\end{eqnarray}
where $\leftidx{^{(4)}}{R}_{ab} = \leftidx{^{(4)}}{R}_{cacb}$ and $\leftidx{^{(4)}}{R} = \leftidx{^{(4)}}{R}_{aa}$
are the Ricci tensors and the Ricci scalar, respectively,
determined by the four-dimensional geometry \eq{4-vier} and the Laplacian operator is defined by
\begin{equation}\label{laplac}
  \Delta \phi \equiv D_a \partial_a \phi = \frac{1}{\sqrt{g}} \partial_\mu (\sqrt{g} g^{\mu\nu} \partial_\nu \phi).
\end{equation}

\section{Representation of Ricci tensors}

The Riemann curvature tensors in \eq{riemann-curvature}, under the group $SO(5)$, correspond
to the tensor product
\begin{equation} \label{yt-riemann}
R_{ABCD} \in \mathbf{10} \otimes \mathbf{10}
= {\tiny  \Yvcentermath1 \yng(1,1)} \otimes {\tiny  \Yvcentermath1 \yng(1,1)}.
\end{equation}
The Clebsch-Gordan decomposition of this tensor product is given by
\begin{equation} \label{yt-riemann-dec}
R_{ABCD} \in \mathbf{10} \otimes \mathbf{10}
= {\tiny  \Yvcentermath1 \yng(1,1)} \otimes {\tiny  \Yvcentermath1 \yng(1,1)}
=  \left(\mathbf{50} = {\tiny  \Yvcentermath1 \yng(2,2)} \right)
 \oplus \left(\mathbf{45}
= {\tiny  \Yvcentermath1 \yng(2,1,1)} \right) \oplus \left( \mathbf{5}
= {\tiny  \Yvcentermath1 \yng(1,1,1,1)} \right).
\end{equation}
The last two representations, $\mathbf{45} \oplus \mathbf{5}$, are removed
by the first Bianchi identity \eq{1st-bianchi}.
In particular, the last one $\mathbf{5}$ corresponds to the five constraints \eq{5-bianchi}.
The Clebsch-Gordan decomposition \eq{yt-riemann-dec} can be further decomposed according
to the symmetry breaking pattern \eq{symm-break} or more precisely Eq. \eq{decom-so5}, as was shown in Sec. 3.
It is straightforward to determine the decomposition of
the Ricci tensors and Ricci scalar in Eqs. \eq{kk-51} and \eq{kk-54} using the result \eq{rdecom-abcd}:
\begin{eqnarray} \label{ricci-decomp}
  && R_{ab} =  e^{\frac{1}{3} \phi} \left\{ (f^{ij}_{(++)} \delta^{ij} + f^{ij}_{(--)} \delta^{ij} ) \delta_{ab}
  + 2 f^{ij}_{(+-)}  \eta^i_{ac} \bar{\eta}^j_{bc}
  - \frac{1}{6} ( \partial_a \phi \partial_b \phi - \Delta \phi \delta_{ab} ) \right. \xx
&& \hspace{2cm} \left. - \frac{\kappa^2}{2} e^\phi \Big( (f^{(+)i} f^{(+)i} + f^{(-)i} f^{(-)i})\delta_{ab} + 2 f^{(+)i} f^{(-)j}
\eta^i_{ac} \bar{\eta}^j_{bc} \Big) \right\}, \\
\label{scalar-decomp}
&& R =  e^{\frac{1}{3} \phi} \left\{ 4 (f^{ij}_{(++)} \delta^{ij} + f^{ij}_{(--)} \delta^{ij} )
  - \frac{1}{6} ( \partial_a \phi \partial_a \phi - 2 \Delta \phi  )
  - \kappa^2 e^\phi (f^{(+)i} f^{(+)i} + f^{(-)i} f^{(-)i}) \right\}. \nonumber
\end{eqnarray}
The energy-momentum tensor in Eq. \eq{em-tensor} takes the form \cite{joy-jhep}
\begin{eqnarray} \label{em-decom}
 T_{ab} &=& e^{\sqrt{3} \kappa \Phi}  F_{ac} F_{bc}
 + \partial_a \Phi \partial_b \Phi
 - \delta_{ab} \Big( \frac{1}{4} e^{\sqrt{3} \kappa \Phi} F_{cd} F_{cd}
  + \frac{1}{2} (\partial_c \Phi)^2 \Big) \xx
  &=& 2 e^{\sqrt{3} \kappa \Phi} f^{(+)i} f^{(-)j} \eta^i_{ac} \bar{\eta}^j_{bc}
  + \partial_a \Phi \partial_b \Phi
 - \frac{1}{2} \delta_{ab} (\partial_c \Phi)^2.
\end{eqnarray}

\end{document}